# Haicheng and Tangshan Earthquakes as Potential Dragon-Kings

**Jiawei LI[1], and Didier SORNETTE[1]**

[1]Institute of Risk Analysis, Prediction and Management (Risks-X), Academy for Advanced Interdisciplinary Studies, Southern University of Science and Technology (SUSTech), Shenzhen, China, 518055.

Corresponding author: Didier Sornette (dsornette@ethz.ch), Jiawei Li (lijw@cea-igp.ac.cn)

## Abstract

The dragon-king earthquake hypothesis proposes that some very large to great earthquakes are not merely the extreme end of the Gutenberg-Richter (GR) frequency-magnitude distribution (FMD), but may reflect shifts into different rupture regimes and/or the action of amplifying mechanisms that enhance their likelihood or size, making them statistical outliers. We develop a data-driven framework to systematically test the dragon-king earthquake hypothesis. Our method combines objective spatial clustering, based on data-adaptive kernel density estimation (KDE) of earthquake spatial densities, with a high-power sequential outlier detection technique. For each identified cluster, we examine the tail of the FMD to identify anomalous events. Candidate dragon-kings are evaluated via robust statistical tests, primarily the max-robust-sum (MRS) test with inward sequential testing. For each observed statistic of the MRS test, we calculate its *p*-value defined as the probability that this statistic could be generated by the null distribution. We apply this framework to seismicity surrounding the 1975 $m_L$ 7.4 ($M_W$ 7.0-7.3) Haicheng and 1976 $m_L$ 7.9 ($M_W$ 7.5-7.6) Tangshan earthquakes. For Haicheng, the mainshock shows a strong dragon-king signature in its pre-mainshock sequence, with *p*-values between 0.003 and 0.01 across a stable range of KDE density thresholds used to define natural seismicity clusters. Post-mainshock and combined sequences yield slightly higher *p*-values (up to 0.09). For Tangshan, the mainshock also exhibits statistically significant dragon-king characteristics in both its pre-mainshock and post-mainshock sequences, with *p*-values ranging from 0.02 to 0.05. The evidence that the Haicheng and Tangshan mainshocks exhibit dragon-king characteristics supports the idea that some large and great earthquakes arise from a nucleation process that may enhance predictability. However, the interpretation of the pre-mainshock dragon-king signals for both Haicheng and Tangshan, especially Tangshan, should be treated with caution, because the limited number of pre-mainshock events may bias the construction of the null hypothesis. Haicheng provides a clear case study, while retrospective analyses suggest that Tangshan might also have been forecasted under more systematic procedures.

**Keywords** Dragon-king, Gutenberg-Richter (GR) law, outlier detection, Haicheng earthquake, Tangshan earthquake

## 1 Introduction

### 1.1 From characteristic earthquakes to dragon-king earthquakes

The characteristic earthquake model proposes that, beyond the broad continuum of earthquake sizes, certain faults or fault segments repeatedly generate earthquakes of nearly identical magnitude and slip (Aki, 1984; Schwartz and Coppersmith, 1984). In seismic hazard analysis, the model is often applied by assuming full rupture to estimate the maximum possible event magnitude for individual faults (e.g., Li et al., 2022). In such areas, the overall small to moderate-size seismicity is expected to follow the Gutenberg-Richter (GR) law, upon which a discrete distribution of large characteristic events is superimposed; some studies further predict a gap in intermediate magnitudes separating these events from the maximum expected event (Ben-Zion, 2008; Schwartz and Coppersmith, 1984; Wesnousky, 1994). Recent observations, however, often contradict the model: ruptures routinely cross segment boundaries and produce larger, more complex earthquakes than expected (Antoine et al., 2025; Peng et al., 2025; Ren et al., 2024), exposing the model's sensitivity to subjective fault segmentation and its limited ability to capture the true complexity of fault systems.

Dragon-king (DK) theory offers a fresh perspective on extreme earthquakes, building upon earlier statistical evidence for deviations from the GR distribution at the largest magnitudes. In particular, Pisarenko and Sornette (2004) rigorously identified a crossover in the GR distribution around moment magnitude $M_W$ 8.1 ± 0.3 when analyzing subduction zones collectively, suggesting a different class of earthquakes beyond this threshold. Drawing on four seismological mechanisms (strong fault coupling, finite-fault rupture effects, geometric complexity of rupture, and runaway instabilities revealed in dynamic simulations), Li et al. (2025a) argued that dragon-king earthquakes can be identified as statistical outliers to the GR law. Notably, several case studies have qualitatively discussed the plausibility of large earthquakes being dragon-kings, such as the 2008 Wenchuan event (Li et al., 2025a; Liu et al., 2025) and the 2024 Noto earthquake (Liu et al., 2024). A "dragon-king" is simultaneously an exceptionally large event ("king") and the product of unique circumstances ("dragon") (Sornette, 2009; Sornette and Ouillon, 2012). Building on earlier work that linked the predictability of catastrophic outliers to their distinctive origins (Sornette, 1999; de S. Cavalcante et al., 2013), dragon-king theory posits that special mechanisms can sporadically magnify events that then become extremes, generating runaway disasters—or, on occasion, extraordinary opportunities—that lie well beyond expectations in null hypothesis. When applied to seismicity, the dragon-king concept substantially overlaps with the long-debated idea of characteristic earthquakes. Both envisage unusually large, partly predictable events that deviate from the usual frequency-magnitude distribution (FMD). Yet the emphasis differs: dragon-kings are outliers (which can span several faults) driven by unique physical feedbacks, whereas characteristic earthquakes are recurrent, fault-segment-specific events of similar size. Together these notions underscore the need to understand, and to test rigorously, the processes that produce the largest earthquakes, with direct implications for forecasting and seismic-hazard assessment.

### 1.2 Key challenges in testing the dragon-king earthquake hypothesis

Seismicity analysis inevitably begins with the choice of spatiotemporal window ranges. This decision largely controls the resulting FMD, and because window boundaries are often set by practical or subjective considerations (Li et al., 2025b), any test for dragon-king outliers must guard against spurious detections introduced by arbitrary windowing. By contrast, in fields such as material science, landslides, finance, or sociology, dragon-king studies typically contend only with the length of the time window, not with the additional spatial dimensions. For instance, laboratory rock-fracture experiments are free from subjective spatial windows and typically produce dragon-king-type acoustic emission bursts. During sustained creep, innumerable microcracks nucleate and coalesce, each releasing detectable energy distributed according to an approximate power law similar to the GR distribution. The final rupture manifests as a single energy burst many orders of magnitude larger than all preceding emissions and is mathematically described by a finite-time singularity (Johansen and Sornette, 2000; Gluzman and Sornette, 2001, 2002; Ide and Sornette, 2002; Sornette, 2009; Lei, 2012; Lei and Ma, 2014). Lei (2012) identified at least two generic pathways to such extremes: (i) a power law acceleration in event rate and moment release, and (ii) hierarchical fracture in which rupture jumps from lower to higher structural tiers created by hierarchical inhomogeneities. These results underscore that, once an appropriate observational window is chosen, or rendered unnecessary, dragon-king events can be isolated and studied with minimal ambiguity.

In contrast, at tectonic scales, the dragon-king earthquake hypothesis, like the characteristic earthquake model, has encountered major empirical hurdles. Conclusions regarding characteristic earthquakes remain controversial and warrant far more rigorous scrutiny and stronger statistical methods than they have used to date (Field and Page, 2011; Ishibe and Shimazaki, 2008, 2012; Jackson and Kagan, 2006; Kagan et al., 2012; Main and Naylor, 2012; Nishenko, 1989, 1991; Page et al., 2011). The GR distribution is traditionally applied as a statistical model across broad aggregated regions, assuming a uniform power law behavior. However, this approach overlooks the profound spatial heterogeneity inherent in fault networks, thermal fluxes, rock properties, and other geophysical factors that vary significantly across the Earth's crust. Aggregating seismic events into a single global GR law is akin to analyzing the distribution of human heights without accounting for influential variables like gender, age, or ethnicity and amounts to mixing disparate elements into a homogenized dataset that obscures critical distinctions. Such aggregation risks masking a wealth of relevant geophysical and seismological features, including the presence of distinct earthquake types like dragon-kings, which deviate from the expected GR trend. Mixing heterogeneous populations can cause several paradoxes and statistical pitfalls as exemplified by the Simpson's Paradox (a trend appearing in several groups reverses when the groups are combined), the ecological fallacy (incorrectly inferring individual-level relationships from aggregate data), the aggregation bias (the act of grouping or aggregating data creates spurious patterns or hides real ones), Lord's Paradox (two statisticians reach opposite conclusions about a treatment effect based on the same data depending on how they adjust for baseline variables) and many others.

By studying the GR distribution within coherent, well-defined regions of seismicity rather than in aggregate, we hypothesize that such a region-specific approach can better capture the localized physics driving earthquake behavior, thereby revealing patterns and anomalies that a broad-scale analysis would otherwise conceal. This shift in focus is essential to unravel the rich, differentiated dynamics of seismicity and enhance our understanding of underlying mechanisms.

The first challenge is thus to define and then identify what could be "coherent well-defined regions of seismicity", if they do exist. In previous works concerned with the detection of characteristic earthquakes, a two-step procedure has generally been adopted: (i) identifying a specific fault and (ii) defining a surrounding region where seismicity is evaluated with the goal to test the hypothesis that a characteristic event could exist for that fault. Both choices are rather arbitrary—the fault segment and the analysis window depend on expert judgment—and the resulting conclusions are highly sensitive to these decisions, undermining reproducibility. A second weakness lies in the statistical tools typically used to flag outliers in the GR distribution. An unusually large event that appears inconsistent with the expected power law tail is often cited as evidence for a characteristic earthquake, yet studies such as Field and Page (2011) and Page et al. (2011) have repeatedly failed to reject the null hypothesis that even the largest events conform to the GR model. This failure to reject the null hypothesis may indicate either that the characteristic earthquake hypothesis lacks support or that the statistical tests employed lack sufficient power to detect deviations, highlighting a critical weakness in their ability to resolve this question, given the lack of consideration on the power of utilized tests.

North China in the mid-1970s provides an ideal natural laboratory for testing the dragon-king hypothesis (Mearns and Sornette, 2021a; 2021b). As an intraplate region, North China has experienced migration of large earthquakes along different faults over the past 2000 years (Liu et al., 2011). Between 1966 and 1976, a sequence of six magnitude 6 to 7 earthquakes occurred in this region. Among them, the $m_L$ 7.4 ($M_W$ 7.0-7.3) Haicheng earthquake of 4 February 1975 was preceded by a dense foreshock swarm (Adams, 1976; Jones et al., 1982; Lei et al., 2024) and was successfully forecasted (Scholz, 1977), enabling mass evacuation despite sub-freezing temperatures (Xu et al., 1982; Wang et al., 2006; Sornette et al., 2021). Seventeen months later, the $m_L$ 7.9 ($M_W$ 7.5-7.6) Tangshan earthquake struck without authoritative warning, causing nearly 240,000 fatalities according to official sources (Chen et al., 1988). Here, we propose to re-examine these two sequences through the dragon-king framework to ask two questions: (1) are these large earthquakes statistically distinguishable from their smaller counterparts? and (2) why one event appeared predictable while the other did not (see however Mearns and Sornette, 2021a)? The present study, focused on detecting dragon-king earthquakes, addresses existing limitations through two key advancements: (1) we introduce a robust method to identify "natural" clusters within seismicity and (2) we employ the most powerful statistical tests available to detect outliers. The first approach resolves the challenge of delineating faults and their associated regions by relying on data-driven clustering rather than arbitrary selections. The second leverages cutting-edge statistical techniques, recognized as the most effective presently known, to detect outliers with greater sensitivity and reliability.

## 2 Framework for detecting dragon-king earthquakes

In order to test the hypothesis of dragon-kings in the context of earthquakes, it is crucial to establish a well-defined spatial domain over which the seismicity is assumed to belong to the same statistical generating process. In absence of a clear procedure to define the domain in which candidate dragon-kings could be present, any identified dragon-king earthquake could be dismissed as an artifact. A robust definition of the spatial domain is therefore a foundational step to ensure the credibility and scientific validity of the hypothesis.

### 2.1 Roadmap for identifying dragon-king earthquakes

The framework we developed for detecting dragon-king earthquakes comprises four main steps, each carefully designed to address this critical challenge:

(1). *Defining the Study Region*: The first step focuses on identifying clusters of seismicity using kernel density estimation (KDE). This technique ensures that the study domain reflects natural groupings of earthquakes rather than arbitrary boundaries. For further analysis, this step is pivotal for identifying a set of events that are likely to be associated with the same seismogenesis. Studying the GR distribution within coherent seismicity regions, rather than in aggregate, is crucial to uncover localized geophysical variations and distinct earthquake types like dragon-kings, which broad-scale analysis obscures.

(2). *Identifying Dragon-King Candidates*: Once the study regions are defined, potential dragon-king earthquake candidates are identified by analyzing the statistical distribution of seismicity in each region separately. This involves examining the tail behavior of the FMD in each region and searching for anomalies that deviate markedly from the expected patterns of the generating process.

(3). *Rigorous Statistical Testing*: The next step applies rigorous statistical methods to individually test whether the identified candidates truly represent dragon-king earthquakes. These tests are specifically designed to ensure that the observed anomalies are not due to random variations or errors in data processing but are statistically significant deviations indicative of unique underlying processes.

(4). *Validation of Dragon-King Events*: If dragon-king earthquakes are detected, the final step involves validating these events through further statistical analysis using block tests. This may also include analyzing the geological and geophysical context and physical mechanisms associated with the events, and comparing them to similar phenomena in other regions or contexts.

By systematically following these steps, the framework ensures that any identified dragon-king earthquake is rooted in a robust, well-defined spatial and statistical context. This approach not only strengthens the validity of the findings but also addresses potential criticisms related to the definition and interpretation of the statistical domains.

## 2.2 Defining the study region using kernel density estimation (KDE)

The framework detects spatial clustering of seismicity using data-adaptive KDE. For a given grid with a spatial resolution of 0.01°, the density $f(\text{Lon, Lat})$ at each grid point (Lon, Lat) is estimated based on the distances to all seismicity observed during the time window from $T_1$ to $T_2$. The density $f(\text{Lon, Lat})$ is calculated using the following formula:

$$f_{[T_1,T_2]}(\text{Lon},\text{Lat}) = \frac{1}{S_{\text{norm}}} \sum_{\substack{i=1 \\ t_i \in [T_1,T_2]}}^{N_{\text{raw}}} \frac{1}{2\pi h_i^2} \exp\left[\alpha m_i - \frac{(\text{Lon}-\text{Lon}_i^{\text{eq}})^2 + (\text{Lat}-\text{Lat}_i^{\text{eq}})^2}{2h_i^2}\right] \quad (1)$$

where ($\text{Lon}_i^{\text{eq}}$, $\text{Lat}_i^{\text{eq}}$) and $m_i$ are the spatial coordinates and magnitudes of the $N_{\text{raw}}$ earthquakes occurring between $T_1$ and $T_2$, and $h_i$ is the adaptive bandwidth determined using the square-root rule based on local density (Abramson, 1982). It is thus different from earthquake to earthquake in Equation (1). This formulation effectively "spreads" the concentrated Dirac-like point mass representing each earthquake into a normalized two-dimensional Gaussian density function with an adaptive bandwidth $h_i$. Beyond the Gaussian kernel, the contribution of each earthquake is further weighted by a magnitude-dependent factor $e^{\alpha m_i}$. This additional weighting reflects the well-established idea in statistical seismology that earthquakes have different capacities to trigger subsequent events, a relationship often modeled as an exponential function of magnitude and commonly referred to as the productivity or fertility law. It is thus natural that, when defining seismicity clusters, earthquakes should contribute according to their typical propensity to generate future earthquakes, thereby shaping the clusters themselves. Densely packed micro-events, while numerous, exert limited influence, whereas larger, less frequent earthquakes appropriately exert a greater impact on the overall seismic density $f(\text{Lon, Lat})$. Importantly, this fertility-based weighting also allows the use of the entire seismic catalog within the spatiotemporal window without requiring an arbitrary magnitude cutoff, preserving the full information content of the seismicity. The normalization factor $S_{\text{norm}}$ ensures that the density function $f(\text{Lon, Lat})$ remains properly normalized. Following earlier studies in China (e.g., Li et al., 2025b), we set the exponent $\alpha = 1$.

Then, we identify clusters as groups of earthquakes located within the boundaries of regions where the function $f$(Lon, Lat) exceeds a specific threshold $f_{th}$. Geometrically, this amounts to define a cluster as a set of events located on the corresponding island defined as a compact spatial domain with "altitudes" $f$(Lon, Lat) above the "sea level" $f_{th}$. For a specific $f_{th}$, there are two categories of seismicity clusters. Class I consists of clusters that remain stable when $f_{th}$ changes within reasonable bounds. These clusters are typically denser. In contrast, Class II includes more diffuse (less dense) clusters whose shapes change significantly with variations in $f_{th}$. There is a prior no optimal value for $f_{th}$, so we propose to identify physically meaningful clusters as those that remain approximately unchanged under reasonable variations in $f_{th}$. To evaluate the robustness of clusters when $f_{th}$ is varied, it is essential to map how the clusters change between two or more different $f_{th}$ values. This requires establishing a clear correspondence between clusters identified at each threshold, enabling us to quantify the changes in their structure or earthquake membership to various clusters. This procedure is justified by the observation that the largest clusters have their boundaries and compositions remaining relatively stable under variations in $f_{th}$.

We thus scan $f_{th}$ from $\Delta f_{th}$ to $N_f \Delta f_{th}$ with a step $\Delta f_{th}$ (e.g., $\Delta f_{th} = 0.001$ and $N_f = 1000$ in the present study). For a given $f_{th}^q$, where $0 < f_{th}^q < 1$ and $q = 1, 2, \ldots, N_f$, let us denote the number of clusters as $C$: this means that the density at each grid point inside each of these clusters is larger than or equal to $f_{th}^q$. Next, we calculate the number of observed events in each cluster. Let us denote $E_q^c$ the number of events in the $c$-th cluster at the $q$-th density threshold, with $c = 1, 2, \ldots, C$. We then rank these values in descending order and assign ranks to the clusters, so that $E_q^{(1)} \geq E_q^{(2)} \geq \ldots \geq E_q^{(C)}$. We then define $\sigma_{(c)}$ as the relative change in the number of earthquakes within the cluster with rank number ($c$), calculated as:

$$\sigma_{(c)} = \frac{\left| E_q^{(c)} - E_{q+1}^{(c)} \right| + \left| E_q^{(c)} - E_{q-1}^{(c)} \right|}{E_q^{(c)}} \quad (2)$$

For stable large clusters, the absolute values of $\sigma_{(c)}$ are small. To quantify the stability of these large clusters, we further define

$$\Phi = \sum_{c=1}^{N_\Phi} \sigma_{(c)} \quad (3)$$

where $N_\Phi$ represents the number of largest clusters considered. This value captures the cumulative relative change in the number of earthquakes for the largest clusters, helping to assess their collective stability with respect to variations in the density threshold $f_{th}$.

2.3 Defining completeness and dragon-king candidate magnitudes

Consider the $c$-th cluster at the $q$-th density threshold. It contains a total of $N_{total} = E_q^c$ earthquakes with magnitudes $MAG_{total} = \{m_1, m_2, \ldots, m_{N_{total}}\}$. Let us first rank these magnitudes in ascending order: $MAG = \{m_{(1)} \leq m_{(2)} \leq \ldots \leq m_{(N_{total})}\}$ (Figure 1). Earthquake magnitudes are well-known in general to obey the GR law. The observed magnitude rank-ordering function, rank($m$), which counts the number of events with magnitude $\geq m$, is the empirical complementary cumulative frequency-magnitude distribution (CCFMD). In the limit where the number $N_{total}$ of earthquakes goes to infinity, the normalized function rank($m$)/$N_{total}$ converges to the theoretical complementary cumulative distribution function (CCDF). Expressing as usual the CCFMD as $\mathrm{CCFMD}(m) = 10^{a - bm}$ where $a$ and $b$ are constants, and $b$ is the slope, typically close to 1, we have the approximate relation

$$\mathrm{rank}(m) \cong \mathrm{CCFMD}(m) = 10^{a-bm} \tag{4}$$

so that $\mathrm{rank}(m = 0)$ corresponds to the expected number $10^a$ of earthquakes of magnitudes larger than or equal to 0. As is well-known, the exponential relation expressing the GR law when earthquake sizes are quantified by their magnitudes becomes a power law when expressed in terms of seismic moments (or energies). Using the method for quantifying power law behavior proposed by Clauset et al. (2009), we obtain a first estimation of the minimum magnitude above which the GR-like power law behavior cannot be rejected. Specifically, for each candidate lower magnitude threshold, we estimate the corresponding GR exponent and compute the Kolmogorov-Smirnov (KS) statistic between the empirical distribution and the fitted power law model. The minimum magnitude is then selected as the smallest threshold for which the power law hypothesis is not rejected at the significance level of 0.05. This choice provides a practical balance: it retains the statistical rigor of the KS-based goodness-of-fit test while partly mitigating its tendency to be overly conservative, which would otherwise exclude too many events from the GR regime. This threshold, denoted as $m_c$, is used as an operational completeness magnitude (Feng et al., 2022; Li et al., 2023; 2025b; 2025d; Wang et al., 2025). It does not necessarily imply that all events above $m_c$ are fully recorded; rather, it defines the magnitude range over which the observed distribution is statistically compatible with the GR law and is therefore suitable for constructing the null hypothesis. More specifically, $m_c$ can be interpreted as the magnitude above which the curvature of the log[CCFMD($m$)], or equivalently the log[rank($m$)], as a function of magnitude approximately vanishes, indicating that the observed distribution enters a nearly linear GR-like regime. This corresponds to a lower magnitude boundary. Next, we select the events with magnitudes larger than $m_c$, forming the set Mag = $\{m_c \leq m_{(1)} \leq m_{(2)} \leq ... \leq m_{(N)}\}$, where $N$ is the number of such events (with $N \leq N_{\mathrm{total}}$) (Figure 1).

We identify an upper magnitude boundary $m_t$ defined as the value above which magnitudes begin to deviate from the GR law, now on the right of the distribution, i.e. for large magnitudes. To quantify the deviation of the rank-ordering function, rank($m$), from the GR law expressed by Equation (4), we use the Anderson-Darling distance $D$ between the empirical distribution and the GR distribution expressed with a first estimate $b^0$ of the slope obtained using the likelihood estimation method (Aki, 1965) combined with 200 bootstrap iterations (Efron, 1979). The Anderson-Darling test emphasizes the tail of the distribution more than other distance metrics (see e.g. Equations (2.35) and (2.36) in Malevergne and Sornette (2006)) and reads:

$$D^2(m_{\mathrm{t}};\mathrm{Mag}) = -N - 2 \sum_{\substack{i=1 \\ m_{(i)} \leq m_t}}^{N} \{\frac{2i}{2N+1} \log_{10}[\frac{\mathrm{rank}(m_{(i)})}{N}] + (1 - \frac{2i}{2N+1}) \log_{10}[\frac{1 - \mathrm{rank}(m_{(i)})}{N}]\} \tag{5}$$

The term $2i/(2N + 1)$ represents by definition the theoretical cumulative distribution function of the $i$th rank, which is compared with $\mathrm{rank}(m_{(i)})/N$ to define the Anderson-Darling distance $D$. The upper magnitude boundary $m_t$ is estimated as the magnitude that minimizes the Anderson-Darling distance $D$ (Figure 1). The $r$ events with magnitudes larger than $m_t$ are considered as dragon-king earthquake candidates (Figure 1).

Next, we incorporate this new upper bound and recalculate the lower magnitude $m_c$ using data from the set MAG = $\{m_{(1)} \leq m_{(2)} \leq ... \leq m_{(N\mathrm{total})}\}$ of magnitudes all smaller than $m_t$ in order to avoid the possible bias introduced by the large events deviating from the GR law. Finally, the set Mag = $\{m_c \leq m_{(1)} \leq m_{(2)} \leq ... \leq m_{(N)}\}$ is then divided into two parts: the $r$ dragon-king earthquake candidates constituting the set $\text{mag}^* = \{m_t \le m_{(N-r+1)} \le m_{(N-r+2)} \le \ldots \le m_{(N)}\}$, and the remaining events that are expected to follow the GR law, corresponding to the magnitude set $\text{mag} = \{m_c \le m_{(1)} \le m_{(2)} \le \ldots \le m_{(N-r)} < m_t\}$. At the end of this step, the slope $b$ of the GR law is recalculated using the set $\text{mag} = \{m_c \le m_{(1)} \le m_{(2)} \le \ldots \le m_{(N-r)} < m_t\}$ and is denoted $b^{\text{mag}}$, again to be fully consistent and avoid using the earthquakes that have been shown to deviate from the candidate GR law both on the left (small events) and on the right (large events) of the distribution.

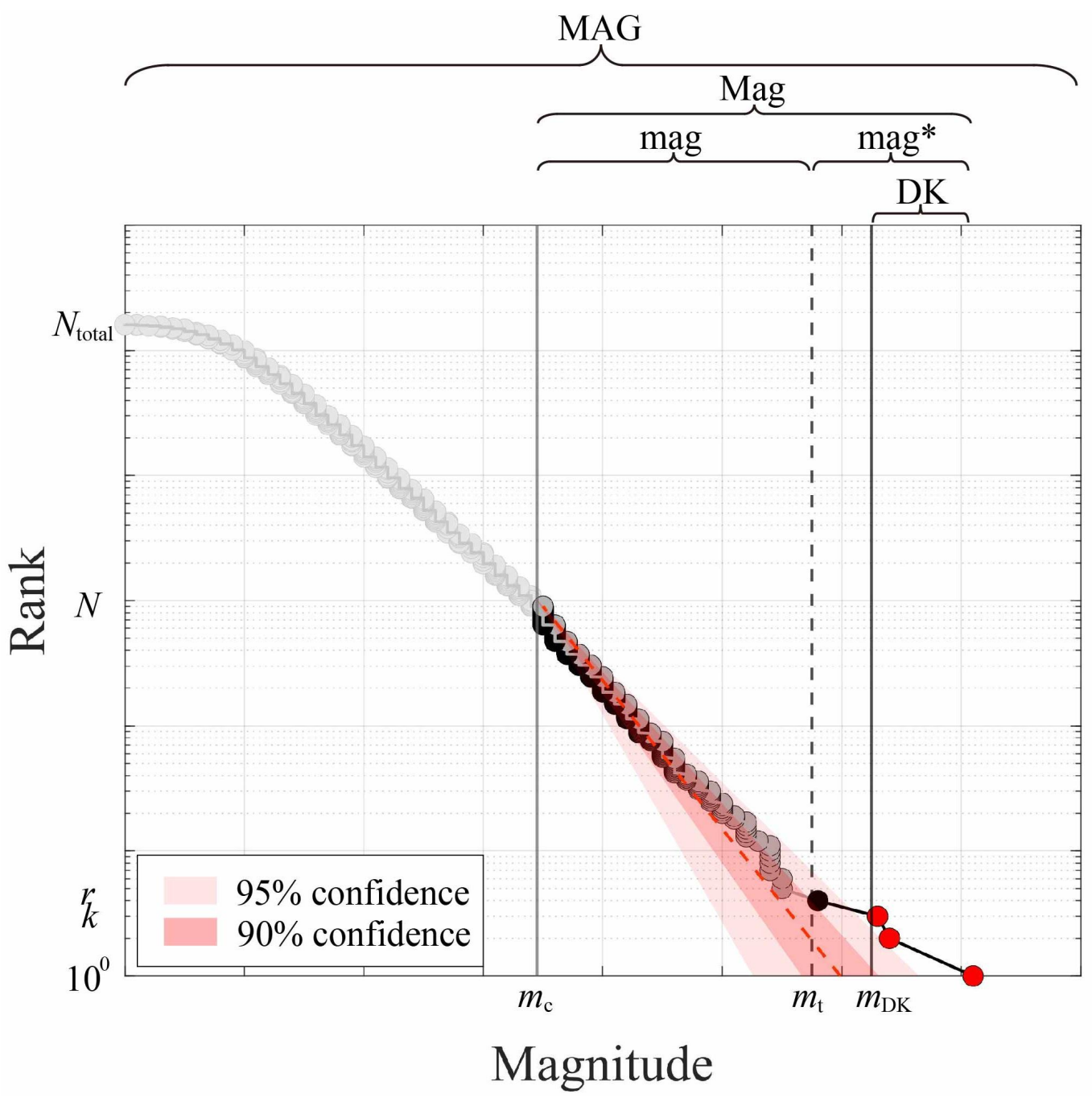

**Figure 1**. Schematic illustration of magnitude rank-ordering plots (non-normalized complementary cumulative frequency-magnitude distribution, CCFMD) for seismicity within a given cluster at a specified density threshold $f_{\text{th}}$. The dragon-king earthquake testing framework developed in the present study partitions the frequency-magnitude distribution (FMD) into three distinct segments using two magnitude thresholds: the completeness magnitude $m_c$ (vertical solid gray line) and the candidate dragon-king transition magnitude $m_t$ (vertical dashed gray line). These segments include the incomplete part (light gray solid circles), the Gutenberg-Richter (GR) part (dark gray solid circles), and the dragon-king candidate part (black solid circles). Dragon-king earthquakes identified at the significance level $\alpha = 0.05$ (red solid circles) are bounded below by the smallest magnitude among them, which defines the dragon-king transition magnitude $m_{\text{DK}}$ (vertical black line). The theoretical GR law, represented as $\text{CCFMD}(m) = 10^{a - bm}$, is fitted to the data between $m_c$ and $m_{\text{DK}}$ and shown as the red dashed line, with 90% and 95% confidence intervals corresponding to significance levels $\alpha = 0.1$ and $\alpha = 0.05$, respectively. Key magnitude sets are defined as: (1) MAG: the raw set of $N_{\text{total}}$ earthquakes in the cluster, $\{m_{(1)} \le m_{(2)} \le \ldots \le m_{(N\text{total})}\}$; (2) Mag: the subset of $N$ events with magnitudes $\ge m_c$, $\{m_c \le m_{(1)} \le m_{(2)} \le \ldots \le m_{(N)}\}$; (3) mag: the subset of $N - r$ events in Mag with magnitudes $< m_t$, used to fit the GR law, $\{m_c \le m_{(1)} \le m_{(2)} \le \ldots \le m_{(N-r)}\}$; (4) $\text{mag}^*$: the set of $r$ candidate dragon-king earthquakes with magnitudes $\ge m_t$, $\{m_t \le m_{(N-r+1)} \le m_{(N-r+2)} \le \ldots \le m_{(N)}\}$; (5) DK: the final set of $k$ dragon-king earthquakes identified from $\text{mag}^*$ via statistical testing, $\{m_{(N-k+1)} \le m_{(N-k+2)} \le \ldots \le m_{(N)}\}$, where $m_{\text{DK}} := m_{(N-k+1)}$. This schematic captures key features that recur throughout the analysis presented in the results below.

### 2.4 Identifying dragon-king earthquakes

Having identified $r$ dragon-king earthquake candidates, we apply statistical outlier (or dragon-king) tests (see Sornette and Wei (2025) for a recent review and extensions of standard outlier detection tests). We adopt the inward sequential testing method with the max-robust-sum (MRS) test statistic (Sornette and Wei, 2025) to identify dragon-king earthquakes from the set $\text{mag}^*$. The MRS test statistic for $m_{(j)}$ is defined as

$$T_{j,N-v}^{\mathrm{MRS}} = \frac{m_{(j)}}{\sum_{i=1}^{N-v} m_{(i)}} \tag{6}$$

Here, we test dragon-king earthquake candidates in descending order, starting with $j = N$ (the largest event), then $j = N$-1, $N$-2, ..., and continuing until the null hypothesis is first not rejected—no later than $j = N$-$r$+1. In the denominator of Equation (6), the sum extends from $i = 1$ to $N$-$v$ with $v \geq r$, ensuring that the reference sample is not contaminated (swamped) by the outliers (Sornette and Wei, 2025). In this study we set $v = r$, a choice that offers a practical compromise between preventing outlier contamination (swamping) and retaining a sufficiently large reference sample for stable estimation (see the Results section for further discussion of the case $v > r$). It amounts to comparing the candidate outlier magnitude $m_{(j)}$ to the sum of the $N$-$r$ smaller magnitude in the set mag. Under the null hypothesis that all magnitudes are distributed according to the GR law, the probability distribution of this MRS test statistic $T_{j,N-r}^{\mathrm{MRS}}$ can be analytically calculated. The empirical value of $T_{j,N-r}^{\mathrm{MRS}}$ then provides the corresponding $p$-value, i.e., the probability that the observed $T_{j,N-r}^{\mathrm{MRS}}$ could be generated by the null distribution (see below for the calculation formula). Using the MRS test statistic with the inward sequential testing method is computationally efficient, reduces the occurrence of false positives, and is often more effective at detecting single or multiple scattered outliers (Sornette and Wei, 2025).

To benchmark the method and obtain a more accurate non-asymptotic determination of the $p$-value given the finiteness of the sample and the empirical statistical fluctuations around the pure GR law, we use the central set mag = $\{m_c \leq m_{(1)} \leq m_{(2)} \leq ... \leq m_{(N-r)}\}$ as a benchmark. We thus generate 10,000 Monte Carlo realizations of the null hypothesis, which follow the GR law with slope $b^{\mathrm{mag}}$ and calculate the MRS test statistic for all simulations. The $p$-value is then defined as

$$p\text{-value} = \frac{1}{N_{\mathrm{sim}}} \sum_{i=1}^{N_{\mathrm{sim}}} I(T_i^{\mathrm{sim}} > T^{\mathrm{obs}}) \tag{7}$$

where $I(.)$ is the indicator function that equals 1 when the condition inside is true and 0 otherwise, $T_i^{\mathrm{sim}}$ denotes the test statistic for the $i$-th simulation, and $T^{\mathrm{obs}}$ is the value from the observed seismicity. Given a significance level $\alpha$ (set at $\alpha = 0.05$ in this study), the critical value, $S_c$, is determined as the (1 - $\alpha$) -percentile of the test statistic distribution. If the MRS test statistic derived from the observed seismicity exceeds $S_c$, the null hypothesis $H_0$ is rejected.

We start with the largest magnitude, $m_{(N)}$ (i.e., $j = N$ in Equation (6)) and we ask whether it could be qualified as a dragon-king earthquake according to the MRS test statistic. If the MRS test statistic for $m_{(N)}$ derived from the observed seismicity exceeds $S_c$, then $m_{(N)}$ is declared a dragon-king earthquake. The test is then repeated for $m_{(N-1)}$, and this process continues until the null hypothesis is not rejected for the first time (inward sequential testing). Thus, the estimated number $k$ of dragon-king earthquakes, with magnitudes greater than the dragon-king transition magnitude $m_{\mathrm{DK}}$ (defined as the minimum magnitude among them), is the number of marginal tests rejected, i.e., DK = $\{m_{\mathrm{DK}} := m_{(N-k+1)} \leq m_{(N-k+2)} \leq ... \leq m_{(N)}\}$.

### 2.5 Validating dragon-king earthquakes

We further quantify the significance of dragon-king earthquakes by conducting a block test (Hawkins, 1980) for the set DK = $\{m_{\mathrm{DK}} := m_{(N-k+1)} \leq m_{(N-k+2)} \leq ... \leq m_{(N)}\}$ determined in the previous step. If $k = 1$, corresponding to the existence of a single dragon-king earthquake in DK, we exclude this earthquake from the magnitude set Mag and recalculate $b$. We then repeat the previous steps using the MRS test statistic. If $k > 1$, corresponding to the existence of multiple dragon-king earthquakes in DK, we similarly exclude these earthquakes from Mag, recalculate $b$, and use the sum-robust-sum (SRS) test statistic defined by:

$$T_{k,N-v}^{\mathrm{SRS}} = \frac{\sum_{i=N-k+1}^{N} m_{(i)}}{\sum_{i=1}^{N-v} m_{(i)}} \tag{8}$$

where, as discussed above, we set $v = r$ to balance sensitivity and swamping. To compute the $p$-value for the null hypothesis $H_0$ of no dragon-kings, we perform Monte Carlo simulations again, generating a large number of realizations based on the parameter $b$ estimated from the set $\{m_c \leq m_{(1)} \leq m_{(2)} \leq ... \leq m_{\mathrm{DK}}\}$, as described above. The $p$-value is determined as the proportion of these realizations where the test statistic exceeds the one calculated from the observed seismicity, as defined in Equation (7). To ensure precision to at least two decimal places, we generate 10,000 Monte Carlo realizations as well.

## 3 Seismicity around Bohai Bay

We use seismicity data from the China Earthquake Networks Center (CENC), covering events that occurred in the Bohai Bay region and its surrounding areas from the year 200 to December 31, 2024 (Figures 2 and 3). The seismicity data is compiled from earthquake catalogs spanning three periods. For the period from 200 to the early 1900s, the earthquake catalog is based on historical Chinese documents, local chronicles, and literary works (Institute of Geophysics, State Seismological Bureau and Institute of Chinese Historical Geography, Fudan University, 1990a; 1990b; 1990c). The earliest recorded earthquakes in Chinese historical records occurred in 1831 BCE (before common era) in the Tai Mountains of Shandong. Modern seismologists can determine the intensity distribution at the epicenter and estimate the location and magnitude of these earthquakes based on such records, which often include information such as collapsed buildings and the distribution of casualties (Institute of Geophysics, State Seismological Bureau and Institute of Chinese Historical Geography, Fudan University, 1990a; 1990b; 1990c; Bureau of Earthquake Damage Prevention, State Seismological Bureau, 1995). Therefore, the catalog used in this study for events prior to the early 1900s includes parameterized data with intensity magnitudes ($m_{\mathrm{I}}$). In this period, there are 60 events in the study region with $m_{\mathrm{I}} \geq 4.8$. As shown in Figure 3(a), there is a notable absence of recorded earthquakes between approximately 400 and 900 CE (common era). Although North China has long been a major political, economic, and military center providing favorable conditions for the preservation of historical earthquake records, Figure 3(a) still reveals a substantial catalog incompleteness prior to 1900.

From the early 1900s to the 1980s, with the advent of analog seismographs, the earthquake catalog for this period was primarily based on measuring paper records of seismic events and calculating magnitudes ($m_{\mathrm{L}}$). After the 1980s, China began constructing a digital seismic network, and vast amounts of digital seismic records were collected. From this point onward, the catalog began to use digital records to measure magnitudes ($m_{\mathrm{L}}$). To maintain consistency with the previous stage, where magnitudes were calculated using analog records, the methodology for magnitude calculation in the digital period continued to follow similar procedures (Bureau of Earthquake Monitoring and Forecasting, China Earthquake

Administration, 2023). The dataset using $m_L$ after the 1900s includes 95,939 events with $m_L \geq 0$, 67,823 events with $m_L \geq 1.0$, 16,867 events with $m_L \geq 2.0$, 3,833 events with $m_L \geq 3.0$, 1,359 events with $m_L \geq 4.0$, 139 events with $m_L \geq 5.0$, 12 events with $m_L \geq 6.0$, 5 events with $m_L \geq 7.0$, and no events with $m_L \geq 8.0$. These numbers indicate that, even after the 1980s, the earthquake catalog for events with $m_L < 3.0$ remained notably incomplete, underscoring limitations in the detection capabilities and coverage of earlier seismic monitoring systems.

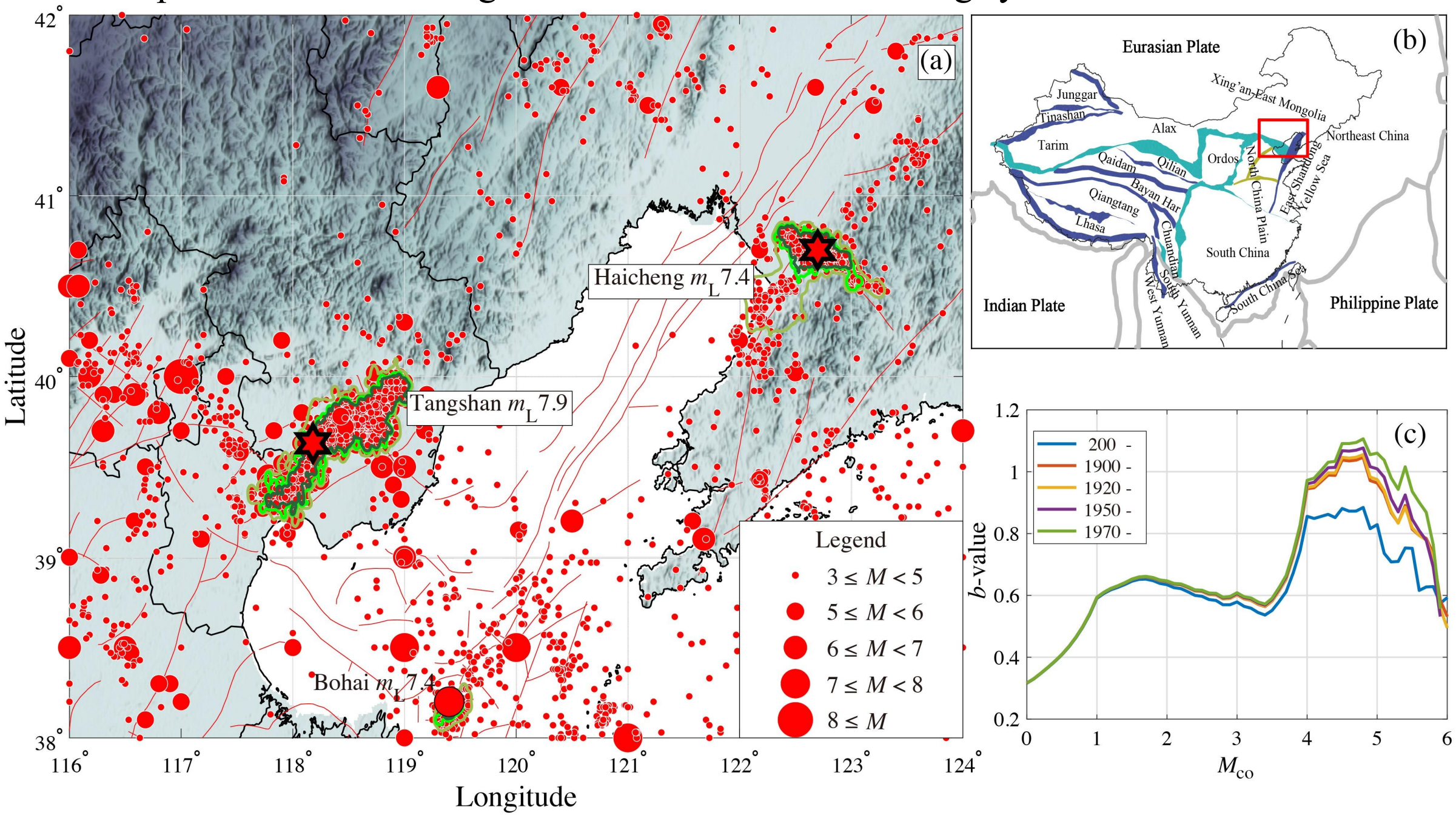

**Figure 2**. (a) Epicenters of earthquakes with $m \geq 3.0$ from the China Earthquake Networks Center (CENC) that occurred from the year 200 to 2024 in the vicinity of Bohai Bay, China. (b) Location of the study region shown on a broader regional map, illustrating the tectonic plates and active tectonic blocks of Chinese mainland (Zhang et al., 2003). (c) Variation of the $b$-value of the GR law as a function of the cut-off magnitude $M_{co}$ for different starting years (200, 1900, 1920, 1950, 1970) up to 2024 within the study region. The methods to estimate the $b$-value are described in Section 2.3. In (a), thin black lines indicate the provincial administrative boundaries of China, while thin red solid lines indicate known mapped faults. The red hexagonal stars mark the $m_L$ 7.4 ($M_W$ 7.0-7.3) Haicheng earthquake on February 4, 1975, and the $m_L$ 7.9 ($M_W$ 7.5-7.6) Tangshan earthquake on July 28, 1976. The red solid circle marks the $m_L$ 7.4 Bohai earthquake on July 18, 1969, which generated a prominent aftershock cluster. Contours delineate the spatial extent of clusters that encompass the three mainshocks at density thresholds $f_{th} = 0.05$ (light green), $f_{th} = 0.15$ (medium green), and $f_{th} = 0.25$ (dark green).

Figure 2(c) shows the variation in the $b$-value of the GR law as a function of the cut-off magnitude $M_{co}$ for different starting years, i.g. 200, 1900, 1920, 1950, and 1970, up to 2024 within the study region. When $M_{co}$ is below 4, the estimated $b$-values across the five starting years show only minor differences. In particular, for the four periods starting after 1900, the $b$-values are nearly identical at $M_{co} < 4$, and they remain within a narrow range of variation (difference < 0.05) across $M_{co}$ values between 4 and 6. This consistency suggests that the seismicity characteristics governing the $b$-value have remained relatively stable over the past century in this region. The two plateaus observed in Figure 2(c) likely reflect two distinct catalog completeness regimes associated with the evolution of seismic monitoring. The first plateau, roughly for $M_{co}$ between 4 and 5.5, corresponds mainly to the mid-1970s period around the Haicheng and Tangshan earthquakes. Although instrumental coverage was still limited, the aftershock sequences of these large events, together with other strong earthquakes such as the offshore $m_L$ 7.4 Bohai earthquake in 1969, contributed a substantial number of events with completeness magnitudes around $m_c \approx 4$. The second plateau, roughly for $M_{co}$ between 1 and 3.5, is more likely associated with the period after 1988, and especially after 2012, when the seismic network improved substantially and many small magnitude earthquakes were recorded. Thus, the two visually separable plateaus can be interpreted as the combined effect of two catalog regimes: one dominated by relatively complete strong earthquake sequences in the 1970s, and the other by improved detection of small events in the modern digital-network period.

To further characterize the earthquake sequences, Figure 4 presents magnitude-time plots for clusters shown in Figure 2 associated with the Haicheng and Tangshan earthquakes delineated using three density thresholds $f_{th}$ = 0.05, 0.15, and 0.25. For the Haicheng sequence, clusters defined by $f_{th}$ = 0.05 and 0.15 both originate with an $m$ 2.8 event on 30 November 1970, whereas the cluster with $f_{th}$ = 0.25 begins with an $m$ 1.7 event on 23 February 1973. In the Tangshan region, the cluster at $f_{th}$ = 0.05 starts with an $m$ 3.4 event on 10 February 1970, while clusters defined by $f_{th}$ = 0.15 and 0.25 both commence with an $m$ 1.6 event on 21 March 1970. Additionally, all three Tangshan clusters include four historical earthquakes prior to 1970 (not displayed in Figure 4): an $m$ 5.0 event in 1562, an $m$ 5.5 event in 1795, an $m$ 5.0 event in 1880, and an $m$ 4.8 event in 1935.

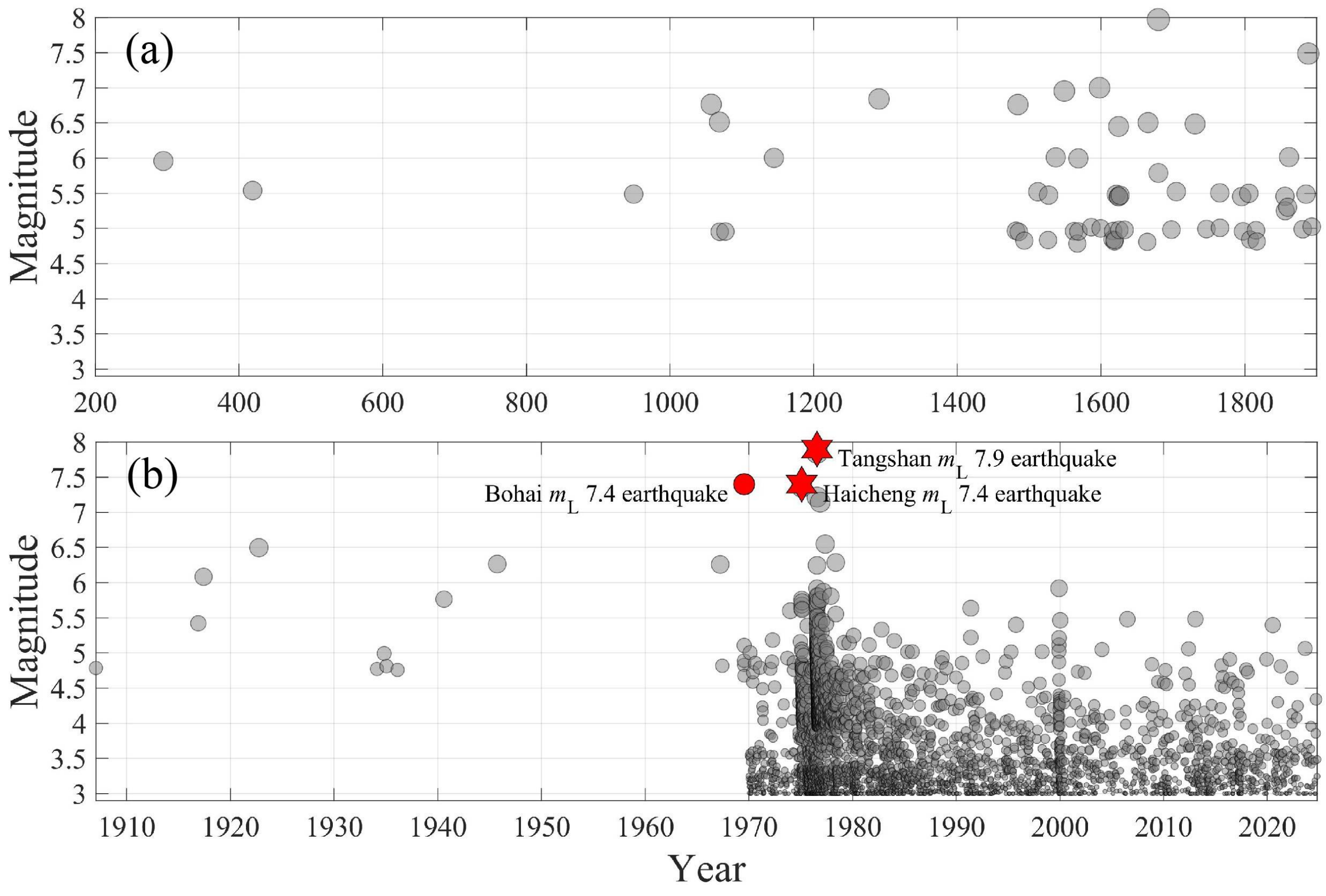

**Figure 3**. (a) The magnitude-time plot of 60 earthquakes with intensity magnitude $m_I \geq 4.8$ that occurred from 200 to 1900 and (b) the magnitude-time plot of approximately 3,900 earthquakes with $m_L \geq 3.0$ that occurred from 1900 to 2024 in the study region.

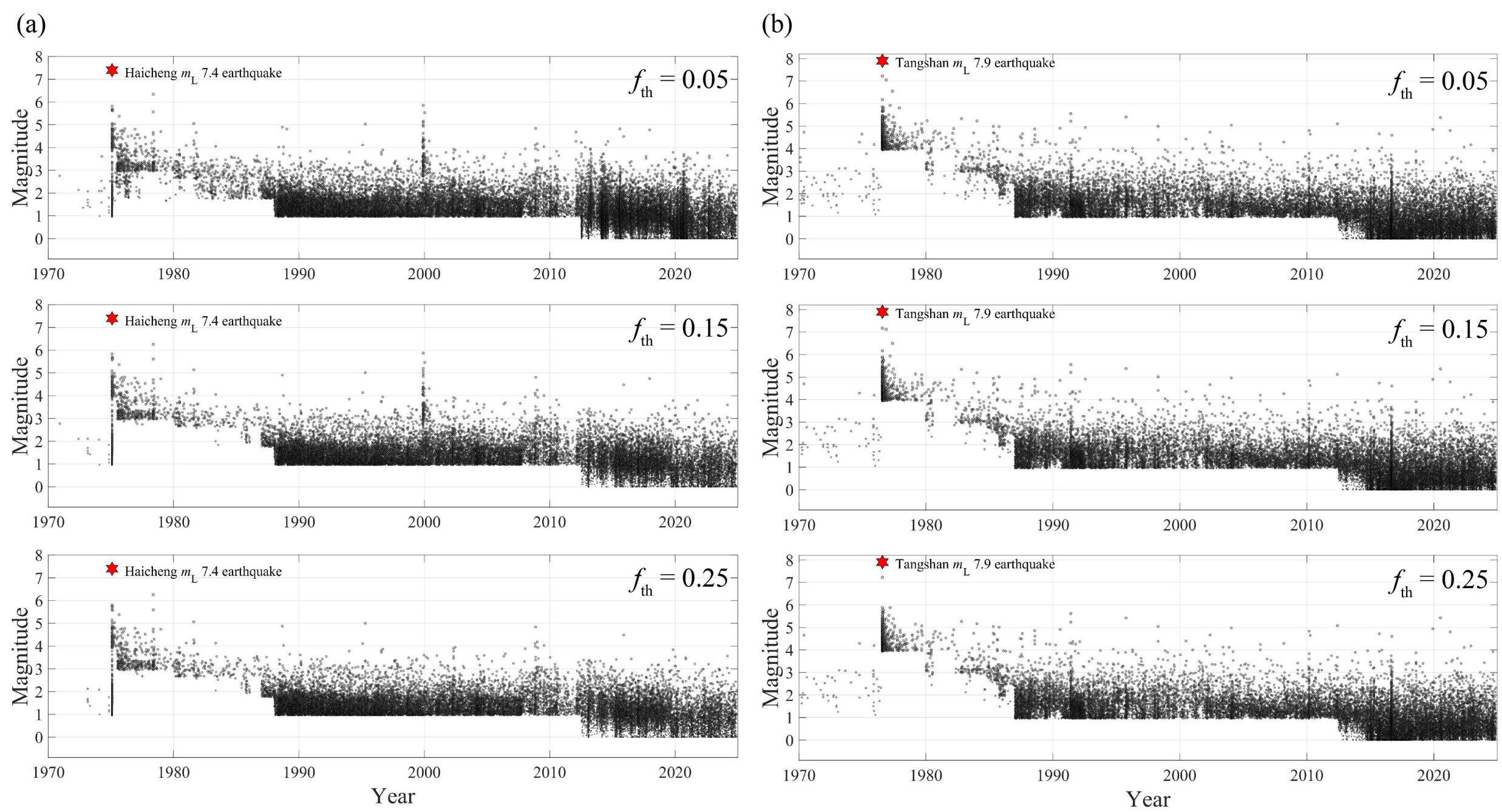

**Figure 4**. Magnitude-time plot of earthquakes within the cluster in Figure 2 encompassing (a) the $m_L$ 7.4 ($M_W$ 7.0-7.3) Haicheng earthquake of 4 February 1975 and (b) the $m_L$ 7.9 ($M_W$ 7.5-7.6) Tangshan earthquake of 28 July 1976, outlined using three different density thresholds $f_{th}$.

## 4 Results

### 4.1 Stability and robustness of identified clusters

Figure S1 presents statistical information on the identified seismicity clusters, including the spatial extent of clusters defined by a given density threshold $f_{th}$, the relative change in the number of earthquakes $\sigma_{(c)}$ as defined in Equation (2) for different clusters, and the rank-order distribution (equivalent to the CCFMD) of the number of earthquakes across the clusters. Figure S1 indicates that, as $f_{th}$ increases, the spatial clustering of seismicity becomes more and more apparent (around $f_{th}$ = 0.05). For larger values of $f_{th}$, clusters containing only a few small-magnitude events undergo significant boundary before eventually disappearing, whereas the boundaries of the larger and denser clusters remain stable and robust with only a few large clusters remaining as $f_{th}$ increases. When $f_{th}$ exceeds 0.3, even the stable large clusters start to lose stability due to the heterogeneous distribution of seismicity within them, eventually splitting into smaller clusters. Figure S2 shows the dependence of Φ defined by Equation (3), namely the sum of $\sigma_{(c)}$ for the top $N_\Phi$ = 1 to 15 spatial cluster regions, as a function of $f_{th}$, demonstrating a trend similar to that shown in Figure S1.

The two main clusters are those associated with the $m_L$ 7.4 Haicheng earthquake on February 4, 1975 (upper right cluster) and the $m_L$ 7.9 Tangshan earthquake on July 28, 1976 (lower left cluster). In addition, there is a clearly identifiable cluster to the south, primarily dominated by the aftershock sequence of the $m_L$ 7.4 Bohai earthquake on July 18, 1969. We have also analyzed this sequence in the Supporting Information; however, due to the poorer observational quality of this offshore event, the reliability of the analysis remains limited. Moreover, several scattered clusters are observed to the west of the Tangshan earthquake, which are mainly composed of historical earthquakes (pre-1950) and their aftershock sequences. Given the even lower completeness of observations in that period, we refrain from further analyzing these clusters in the present study. Figure 2(a) also shows contours that bound the clusters containing the Bohai, Haicheng, and Tangshan mainshocks at density thresholds $f_{th}$ = 0.05 (light green), $f_{th}$ = 0.15 (medium green), and $f_{th}$ = 0.25 (dark green).

4.2 Were the Haicheng and Tangshan earthquakes dragon-king events?

We examine the two main clusters identified by our procedure and the stability condition demonstrated in Figures S1 & S2 to test for the possible existence of dragon-kings. As already mentioned, these two clusters contain respectively the $m_L$ 7.4 Haicheng earthquake on February 4, 1975 (top right cluster), and the $m_L$ 7.9 Tangshan earthquake on July 28, 1976 (bottom left cluster). Testing for the existence of dragon-kings in these two clusters amount to ask if these two main shocks are indeed outliers compared with the distribution of all the earthquakes in their respective cluster.

In order to test for robustness, in addition to examining the outliers for all the events of these two clusters, we also segregate each cluster into two subsets: pre-mainshock events (defined as those occurring prior to the mainshock) and post-mainshock events (defined as those occurring after the mainshock). Figures S3 and S4 present the magnitude rank-ordering plots (non-normalized CCFMD) for the events in each cluster. We also show the distributions for the sequences of pre-mainshock and post-mainshock of the Haicheng and Tangshan earthquakes, respectively, for different density thresholds $f_{th}$. To facilitate reader interpretation, we have extracted the magnitude rank-ordering plots for $f_{th}$ = 0.05, 0.15, and 0.25, whose cluster boundaries and magnitude-time sequences are shown in Figures 2 and 4, respectively, and compiled them into Figures 5 and 6.

The dragon-king earthquake testing framework developed in Section 2 divides the FMD into three distinct regions based on the completeness magnitude $m_c$ and the dragon-king earthquake candidate transition magnitude $m_t$: the incomplete part, the GR part, and the dragon-king candidate part. When dragon-king earthquakes are identified at a significance level of $\alpha$ = 0.05, their lower bound is defined by the minimum magnitude among them, termed the dragon-king transition magnitude $m_{DK}$. Regardless of whether the Haicheng and Tangshan earthquakes are identified as dragon-king events, their $p$-values as potential dragon-king earthquakes, calculated using the MRS test statistic based on the other events in the sequence, are shown in parentheses in the legend in Figures 5, 6, S3 and S4.

We further present in Figures 7 and 8 the $p$-values of the Haicheng and Tangshan earthquakes considered as potential dragon-king earthquakes, plotted as a function of $f_{th}$. In fact, the choice of the significance level $\alpha$ is of secondary importance, as readers can assess the dragon-king earthquake hypothesis using their own $p$-value threshold, informed by their prior beliefs. The red solid and dashed lines in Figures 7 and 8 represent the significance levels ($\alpha$ = 0.1 and $\alpha$ = 0.05, respectively) used in previous studies. In the white shaded area (for $f_{th}$ approximately between 0.05 and 0.25), we confirm that the largest seismicity clusters remain stable and robust. When varying $f_{th}$, we observe small $p$-values that collectively support rejecting the null hypothesis of no outlier (or no dragon-king earthquake).

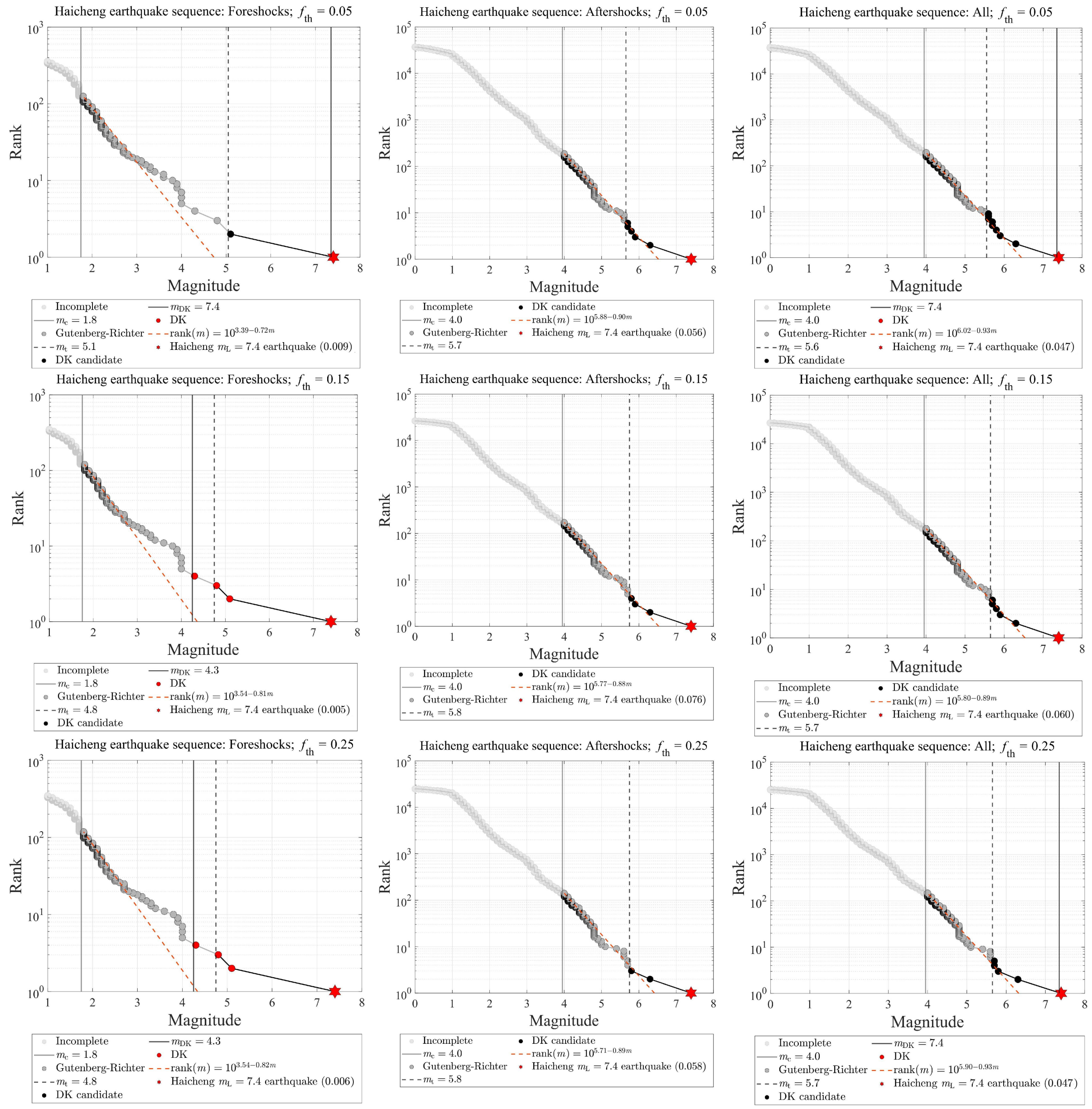

**Figure 5**. Magnitude rank-ordering plots (non-normalized CCFMD) for the sequences of pre-mainshock (left column), post-mainshock (middle column), and combined pre-mainshock and post-mainshock (right column) sequences of the $m_L$ 7.4 ($M_W$ 7.0-7.3) Haicheng earthquake on February 4, 1975, for three density thresholds $f_{th}$ = 0.05 (top line), 0.15 (middle line), and 0.25 (bottom line). The $p$-value of the $m_L$ ($M_W$ 7.0-7.3) 7.4 Haicheng earthquake as a potential dragon-king earthquake, obtained with the max-robust-sum (MRS) test statistic using the other events in the sequence, is displayed in the parentheses of the legend. Symbol conventions are the same as in Figure 1.

For $f_{th}$ ranging from 0.05 to 0.25, the $p$-value of the $m_L$ 7.4 Haicheng earthquake as a potential dragon-king ranges between 0.003 and 0.01 in its pre-mainshock sequence, indicating a notable outlier characteristic consistent with the dragon-king hypothesis. In the post-mainshock and combined pre-mainshock and post-mainshock sequences, the $p$-values range from 0.04 to 0.09, without exceeding 0.1. For the $m_L$ 7.9 Tangshan earthquake, the mainshock also exhibits statistically significant dragon-king characteristics in both its pre-mainshock and post-mainshock sequences, with $p$-values ranging from 0.02 to 0.05. However, compared with Haicheng, the pre-mainshock dragon-king signal of the Tangshan mainshock is less pronounced and less statistically compelling. This contrast suggests that both mainshocks can be identified as dragon-king events, but their temporal expressions differ: Haicheng shows a stronger pre-mainshock dragon-king signature, whereas Tangshan exhibits a more comparable or stronger dragon-king signature in its post-mainshock sequence. If $v$ in Equation (8) is set larger than $r$ ($v > r$), for instance to the rank of the $m_L$ 4.1 event in Tangshan's pre-mainshock sequence in Figure 6, the $p$-value becomes slightly smaller because the reference sample is further protected from possible swamping by moderate anomalous events. The cost, however, is a smaller reference sample and an additional tuning parameter, which may reduce stability and reproducibility. We therefore keep $v = r$ in the main analysis as a conservative compromise between sensitivity and robustness.

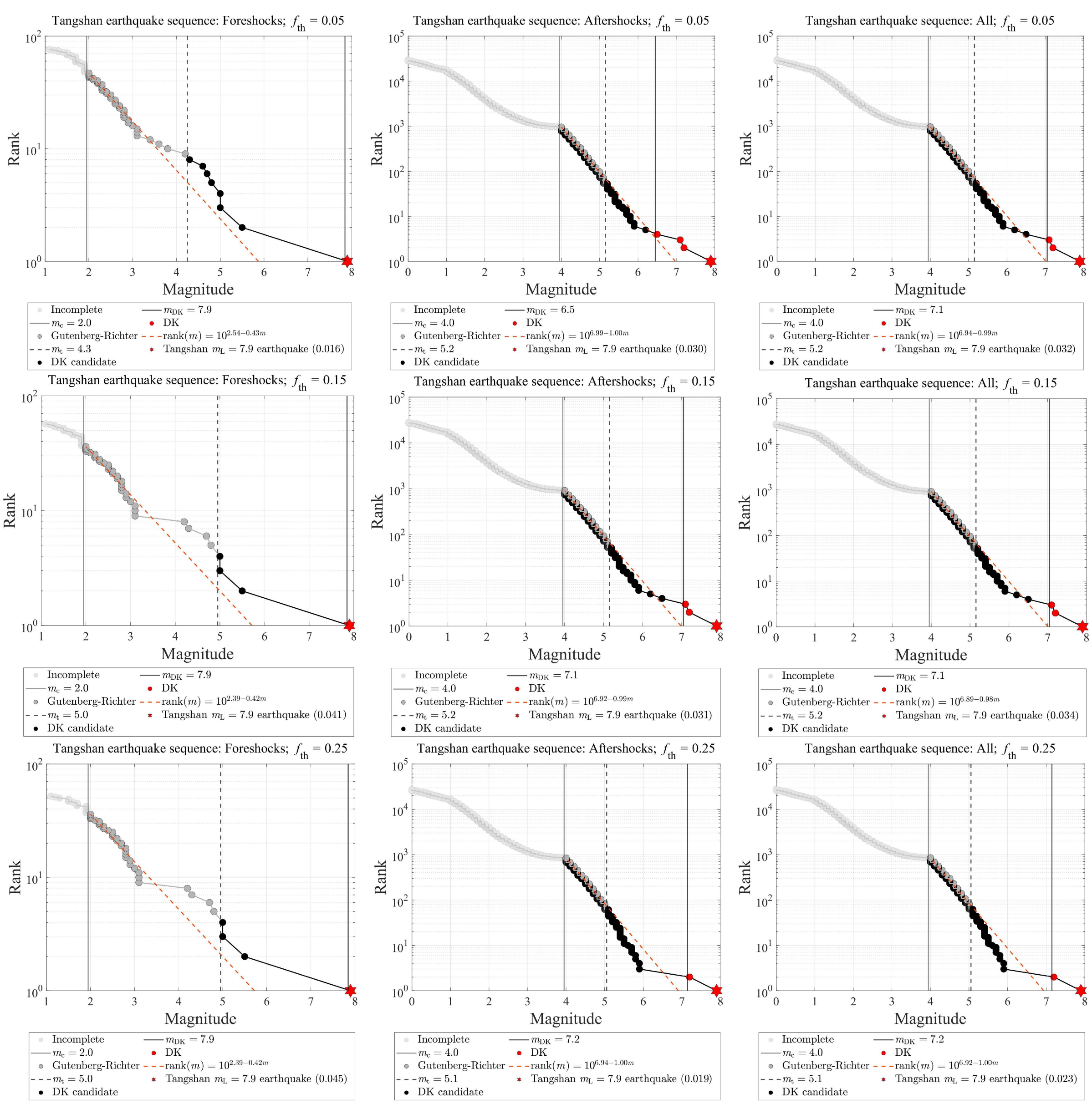

**Figure 6**. Same as Figures 5 for the $m_L$ 7.9 ($M_W$ 7.5-7.6) Tangshan earthquake on July 28, 1976.

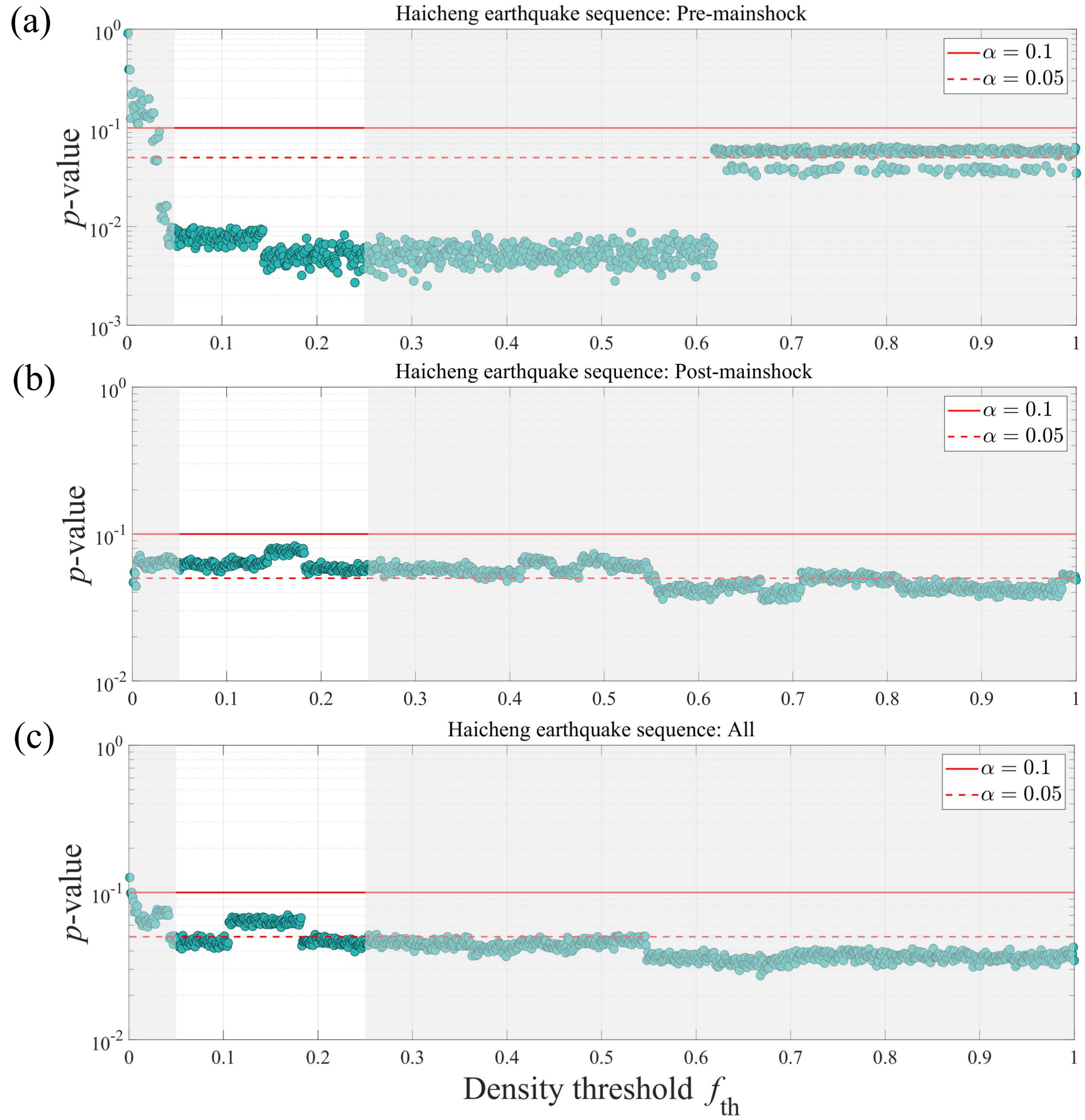

**Figure 7.** For the sequences of (a) pre-mainshock, (b) post-mainshock, and (c) combined pre-mainshock and post-mainshock sequences of the $m_L$ 7.4 ($M_W$ 7.0-7.3) Haicheng earthquake on February 4, 1975, the *p*-value of the $m_L$ 7.4 ($M_W$ 7.0-7.3) Haicheng earthquake as a potential dragon-king earthquake, obtained with the MRS test statistic using the other events in the sequence, is plotted as function of the density threshold $f_{th}$. The red solid and dashed lines represent the confidence levels ($\alpha = 0.1$ and $\alpha = 0.05$, respectively) commonly used in previous studies. In the white shaded area (for $f_{th}$ approximately between 0.05 and 0.25), we checked that the largest clusters of seismicity are stable and robust.

## 5 Discussion

### 5.1 Sensitivity analysis of catalog incompleteness for pre-mainshock sequences

A common concern in analyzing historical earthquake sequences is the potential incompleteness of early seismic catalogs, which could bias statistical estimates of the GR law and, consequently, the results of dragon-king tests. To evaluate the robustness of our findings against this issue, we conducted a sensitivity analysis focused on the pre-mainshock sequences of the Haicheng and Tangshan earthquakes, using the clusters defined by density thresholds $f_{th}$ = 0.05, 0.15, and 0.25.

For the Haicheng earthquake, the pre-mainshock clusters at all three density thresholds contain no events prior to 1970. The earliest events in these clusters are an *m* 2.8 earthquake on 30 November 1970 (for $f_{th}$ = 0.05 and 0.15) and an *m* 1.7 earthquake on 23 February 1973 (for $f_{th}$ = 0.25). Consequently, any potential incompleteness in the catalog before 1900, or even 1970, has no bearing on our analysis of the Haicheng pre-mainshock sequence, as the relevant period for this study is well within the era of complete instrumental recording.

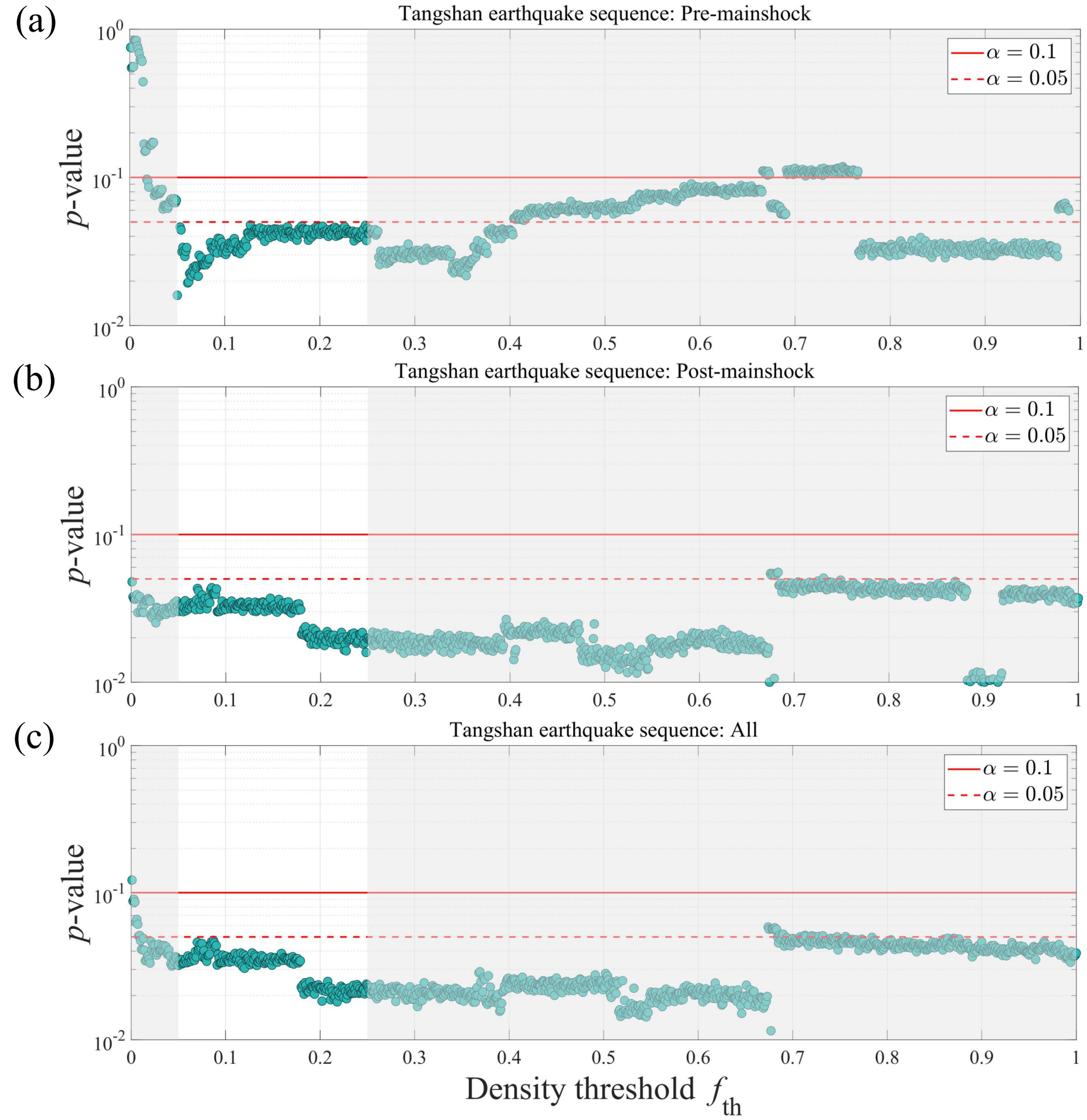

**Figure 8.** Same as Figure 7 for the $m_L$ 7.9 ($M_W$ 7.5-7.6) Tangshan earthquake on July 28, 1976.

The case of the Tangshan earthquake is more complex, as all three clusters include four historical earthquakes prior to 1970: an *m* 5.0 event in 1562, an *m* 5.5 in 1795, an *m* 5.0 in 1880, and an *m* 4.8 in 1935. The completeness for such early historical events is undoubtedly uncertain. However, the impact of this incompleteness on our dragon-king test results is likely negligible. As shown in Figure 2(c), the estimated *b*-value for the region remains remarkably stable (varying by less than 0.02) for cut-off magnitudes $M_{co}$ between 2.0 and 4.0, regardless of whether the catalog starts in 200, 1900, or 1970. This stability indicates that the fundamental seismicity pattern captured by the GR law is well-constrained by the more complete modern data. Since the dragon-king test primarily relies on the fit and deviation from this stable GR distribution derived from the bulk of the data (dominated by modern, complete records), the inclusion or exclusion of a few pre-1900 events is unlikely to significantly alter the calculated *p*-value. Therefore, we conclude that, within our analytical framework, the potential incompleteness of the early catalog does not materially affect the statistical significance of our main findings for both the Haicheng and Tangshan sequences.

5.2 Physical interpretation of dragon-king earthquakes

The small $p$-values associated with the Haicheng and Tangshan earthquakes are direct statistical outcomes of our FMD analysis. In our framework, this result qualifies them as dragon-kings: distinctive earthquakes that stand apart from the rest of the regional seismic population. Their statistical anomaly reflects an excess occurrence frequency far higher than predicted by the GR distribution, i.e., higher than they "should" be if they followed the same scale-free law as ordinary, self-arresting earthquakes (Sornette et al., 2024). According to Li et al. (2025a), strong fault coupling, complex fault rupture structures and unstable run-away rupture represent three key classes of mechanisms that can sporadically magnify earthquake size beyond the expectations of the GR law. First, when the coupling strength of a fault system (e.g., effective normal stress and stiffness) is sufficiently large, individual fault segments can become strongly synchronized so that rupture organizes into system-wide extreme events, rather than remaining confined to self-arresting asperities. Theoretical and observational studies in various tectonic settings show that highly coupled fault patches correlate with the rupture of very large earthquakes (Scholz, 2010). Second, paleoseismological and theoretical work demonstrates that multi-segment cascading ruptures across geometrically complex fault networks can lead to anomalously large ruptures, as envisaged in Vere-Jones' branching rupture model (Vere-Jones, 1977) and subsequent work on complex strike-slip ruptures. When asperities are embedded within a highly complex fault geometry, they do not change the fundamental scaling $P \propto a^2$ that radiated power is always proportional to the square of slip acceleration (Johansen and Sornette, 1999). What they do change is the acceleration itself. Geometric complexity forces the rupture front through repeated episodes of strong deceleration and re-acceleration as it negotiates barriers, jogs, and branching structures. These intermittent bursts of extreme acceleration can be far larger than those produced on a smoother fault. As a result, the $P \propto a^2$ scaling amplifies these peaks disproportionately, creating rare, high-energy outliers in the radiated power. In this way, complex fault geometry becomes a natural generator of rupture dragon-kings: extreme events arising from the constructive interference of multiple localized accelerations rather than from a simple extrapolation of ordinary rupture dynamics. Third, numerical simulations and dynamic rupture theory highlight a distinction between self-arresting earthquakes and unstable run-away ruptures (including sub-shear and supershear ruptures): in run-away events, positive feedback between stress release and rupture propagation overcomes frictional arrest, requiring additional dissipation through seismic wave radiation to achieve dynamic balance (Wen et al., 2018; Wei et al., 2021). In statistical terms, self-arresting earthquakes populate the GR bulk, whereas run-away events appear as dragon-king outliers.

Against this theoretical backdrop, the Haicheng earthquake offers a textbook example of a dragon-king event shaped by a long maturation process involving fluids and progressive fault weakening. Numerous observational and re-assessment studies have documented rich precursory phenomena preceding the 4 February 1975 Haicheng mainshock, including a pronounced foreshock sequence, ground deformation, long-lived tilt anomalies, fluctuations in groundwater level, radon emission anomalies, and reports of anomalous well behavior and animal response (Jones et al, 1982; Molnar, 1981; Wang et al., 2006; Lei et al., 2024). The foreshocks migrated toward the eventual rupture area and concentrated along fluid-rich fault zones, while geodetic and hydrological observations indicated progressive dilatancy and fluid diffusion (e.g., Scholz, 2019). These features are consistent with a maturation or critical-transition process, in which stress accumulation, microcracking, and fluid redistribution gradually weaken the seismogenic fault, bringing it closer to an unstable state. In this context, the low $p$-values (0.003–0.01) obtained for the Haicheng mainshock in its pre-mainshock sequence quantitatively capture the statistical manifestation of such physical amplification processes: the event attains a size far larger than expected from the background GR scaling, standing out as a genuine dragon-king rather than a mere tail fluctuation, because the system has been driven into a highly organized, strongly correlated state. The convergence between statistical anomaly and multiple lines of precursory evidence reinforces the interpretation of Haicheng as a genuine dragon-king event, and illustrates how statistical diagnostics can reveal underlying physical organisation in earthquake preparation.

The Tangshan earthquake, in contrast, exhibits a different dragon-king signature that is more strongly expressed in its source complexity and regional context than in the richness of its immediate precursory sequence (Mearns and Sornette, 2021b). It is well known that Tangshan did not display the same level of short-term precursory activity as Haicheng, even though regional anomalies in the broader Beijing-Tianjin-Tangshan area were reported to continue evolving after 1975 (Mearns and Sornette, 2021a). Source process studies of the 1976 Tangshan sequence show that the mainshock involved right-lateral slip on a NNE-striking fault combined with additional ruptures on adjacent segments, and that the crustal deformation pattern is best explained by a multi-segment, cascading rupture within a rhombic block structure on the northern margin of the North China Basin (Nábělek et al, 1987; Huang and Yeh, 1997). Recent high-resolution imaging of the Tangshan fault zone further reveals a complex fault geometry with multiple strands and step-overs, which provides natural asperities and geometric barriers that can promote intermittent acceleration and deceleration of rupture (Zhang et al, 2022). These findings are consistent with the "complex fault rupture structure" mechanism proposed in Li et al. (2025a): a strongly coupled, geometrically intricate fault network in which a single event can cascade across several segments, producing a rupture that is anomalously large relative to the background GR scaling.

From the perspective of rupture dynamics, Tangshan can thus be interpreted as a candidate unstable run-away rupture in a strongly coupled intracontinental fault system. In such a system, positive feedback between stress release and rupture growth allows the rupture to propagate across multiple segments before arresting mechanisms (friction and radiative damping) can halt it, in contrast to smaller, self-arresting events that populate the GR bulk. This physical picture is reflected in our statistical results: the Tangshan mainshock exhibits statistically significant dragon-king characteristics in both its pre-mainshock and post-mainshock sequences, with $p$-values ranging from 0.02 to 0.05. However, compared with Haicheng, whose pre-mainshock $p$-values range from 0.003 to 0.01, the pre-mainshock dragon-king signal of Tangshan is less pronounced. In other words, both mainshocks can be identified as dragon-king earthquakes, but their temporal expressions differ. Haicheng shows a clearer and more compelling pre-mainshock dragon-king signature, consistent with its richer foreshock activity and multiple reported precursory anomalies. Tangshan also shows a statistically significant pre-mainshock signal, but this signal is weaker and less clearly expressed, whereas its post-mainshock and combined sequences further reinforce its outlier character within a strongly coupled, geometrically complex fault network.

These results indicate that the dragon-king property should not be viewed as a purely intrinsic label independent of the comparison set. Rather, its statistical expression depends on the temporal window used to construct the null hypothesis. A pre-mainshock sequence is the most relevant window for short-term forecasting, because it tests whether the mainshock already stands out relative to the seismicity available before rupture. A post-mainshock or combined sequence, in contrast, is more informative for characterizing the full rupture process, the subsequent stress redistribution, and long-term hazard implications. In this sense, the dragon-king framework has distinct meanings for short-term forecasting and long-term hazard assessment. For Haicheng, the strong pre-mainshock signal suggests a mature preparation process that produced relatively clear statistical precursors. For Tangshan, the significant but weaker pre-mainshock signal suggests that precursory information may have existed, but was more diffuse, less robust, or more difficult to interpret operationally. Thus, the framework does not imply deterministic prediction of all extreme earthquakes; rather, it provides a conditional and probabilistic diagnostic whose forecasting value depends on whether the preparation process generates a sufficiently clear and statistically stable precursor sequence.

It is also important to recognize the statistical limitation of the pre-mainshock analyses. The number of pre-mainshock events is limited for both earthquakes, especially for Tangshan, which may affect the estimation of the GR background and the construction of the null hypothesis. As a result, the *p*-values obtained from pre-mainshock sequences may be subject to stronger statistical bias than those obtained from longer post-mainshock or combined sequences. This limitation does not invalidate the detection of dragon-king characteristics, but it calls for caution when interpreting the relative strength of the pre-mainshock signals and highlights the need for further validation using better-instrumented earthquake sequences.

Taken together, the Haicheng and Tangshan earthquakes demonstrate that dragon-king behavior in seismicity can arise through different physical pathways: (i) a precursor-rich maturation process with strong fluid involvement and progressive weakening of the fault, as in Haicheng; and (ii) a strongly coupled, geometrically complex, potentially run-away rupture in which cascading across multiple segments produces an anomalously large event, as in Tangshan. Both pathways are encompassed by the dragon-king theory, which interprets such events as outliers generated by mechanisms distinct from the bulk of self-arresting earthquakes. Our framework therefore not only provides a quantitative statistical basis for identifying these events as dragon-kings, but also offers a consistent physical interpretation that links their outlier status to strong coupling, complex rupture geometry, and unstable rupture dynamics. This integrated statistical-physical perspective is essential for understanding why some extreme earthquakes may leave strong, detectable precursor signals, while others only reveal their dragon-king nature through their exceptional rupture characteristics and regional impact.

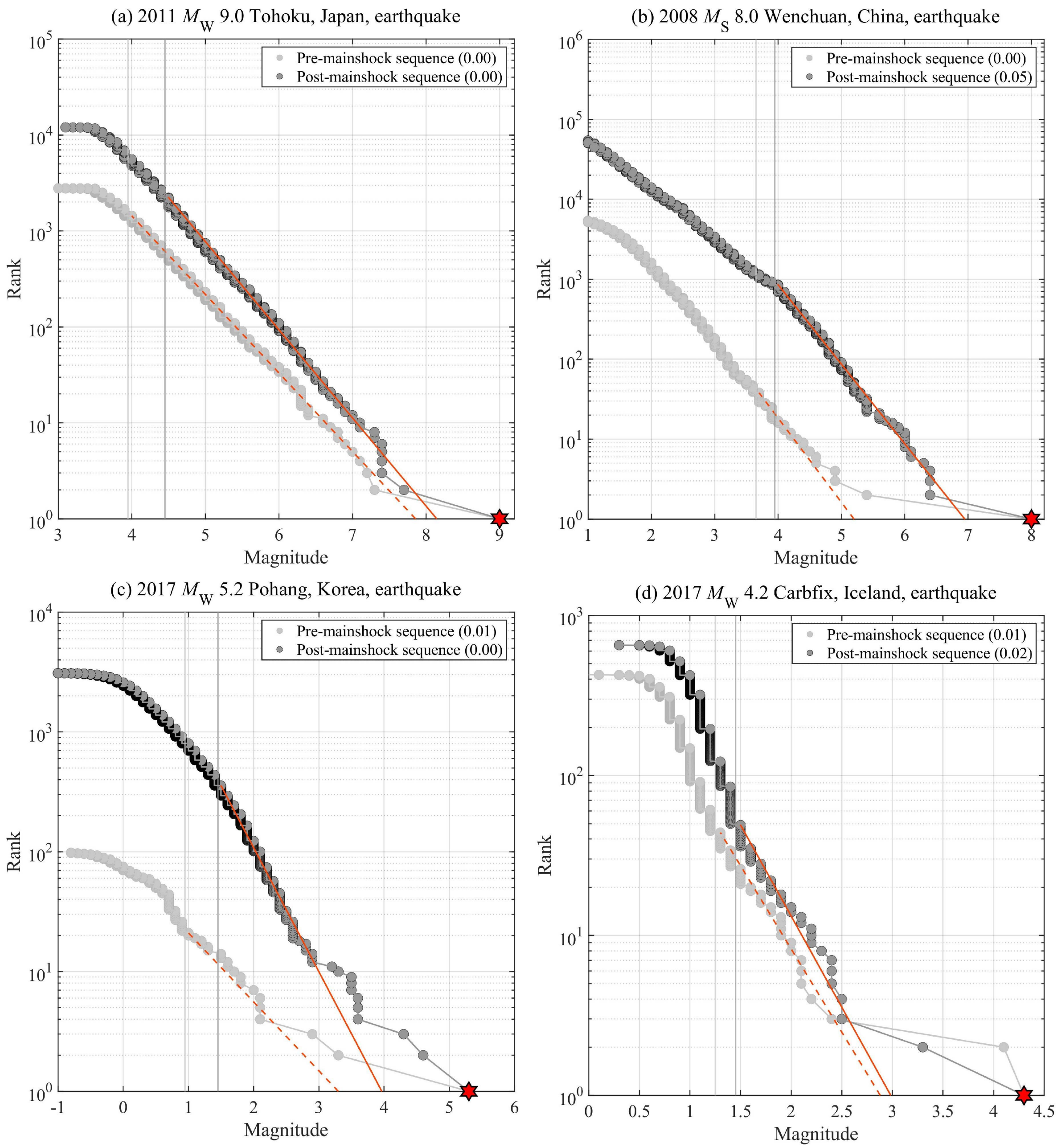

**Figure 9**. Magnitude rank-ordering plots (non-normalized CCFMD) for the sequences of pre-mainshock and post-mainshock of (a) the $M_W$ 9.0 Tohoku, Japan earthquake on March 11, 2011, (b) the $M_S$ 8.0 Wenchuan, China earthquake on May 12, 2008, (c) the $M_W$ 5.2 Pohang, Korea earthquake on November 15, 2017, and (d) the $M_W$ 4.2 Carbfix, Iceland earthquake on December 23, 2017. Panels (a) and (b) represent natural earthquakes. For these, the spatial boundaries for selecting the pre-mainshock and post-mainshock sequences were defined with reference to the aftershock-active regions following the mainshocks. The temporal windows span from 1997 to 2025 for the Tohoku earthquake and from 1970 to 2025 for the Wenchuan earthquake. Panels (c) and (d) represent induced seismicity. The event in (c) is associated with an Enhanced Geothermal System (EGS) in Pohang, Korea. The event in (d) is related to geothermal energy production and $CO_2$ sequestration at the Carbfix site in Iceland. For these induced earthquakes, the spatial and temporal ranges of the seismicity are centered around the respective injection sites and the commencement of injection activities, as provided by the data sources (see Open Research). The red dashed and solid lines are the Gutenberg-Richter relationships fitted to the sequences of pre-mainshock and post-mainshock, respectively. The numbers in parentheses in the legend are the $p$-values of the mainshock as a dragon-king earthquake.

5.3 Towards a broader context: potential dragon-king earthquakes worldwide

In this section, we offer a preliminary exploration of four additional earthquakes from different regions. Our goal is not to repeat the detailed analyses performed for the Haicheng and Tangshan cases, but to highlight similarities that may hint at a broader occurrence of dragon-king behavior. By examining how their GR relations deviate before and after the mainshock, and comparing these patterns with those found in our established dragon-king examples, we outline suggestive evidence that such extreme, structurally amplified events may not be isolated anomalies. These observations are intended as motivation for future, rigorous testing rather than as definitive claims.

As shown in Figure 9, the magnitude rank-ordering plots for the 2011 $M_W$ 9.0 Tohoku (Japan), 2008 $M_S$ 8.0 Wenchuan (China), 2017 $M_W$ 5.2 Pohang (Korea), and 2017 $M_W$ 4.2 Carbfix (Iceland) earthquakes almost all exhibit clear deviations from the GR scaling in both their pre- and post-mainshock sequences, with $p$-values ranging from 0 to 0.02, except for the 2008 Wenchuan mainshock relative to its post-mainshock sequence ($p$-value = 0.05). Notably, the 2011 Tohoku mainshock relative to its pre- and post-mainshock sequences, the 2008 Wenchuan mainshock relative to its pre-mainshock sequence, and 2017 Pohang (Korea) mainshock relative to its post-mainshock sequence show very robust outlier signals ($p$-value = 0). These systematic departures, particularly in the largest events, are consistent with the characteristic signature of dragon-king.

Given computational constraints and time limitations, we focused on a streamlined implementation and did not include the first step of our framework. For the 2011 Tohoku and 2008 Wenchuan earthquakes, we directly adopted the aftershock-active regions following the mainshocks. For the two induced earthquakes (2017 Pohang and 2017 Carbfix), we used samples directly around the respective injection sites and the commencement of injection activities. In other words, the present analysis corresponds to a specific single cluster case, analogous to considering one realisation at a fixed density threshold $f_{th}$ in Figures 7 and 8. These preliminary observations hint that dragon-king behavior may not be limited to North China but could represent a more universal feature of extreme earthquakes across different tectonic and triggering environments. This qualitative assessment must be followed by more systematic and rigorous statistical testing across larger, globally distributed datasets. Future work should aim to apply the full dragon-king testing framework, incorporating robust clustering and powerful outlier detection, to these and other significant sequences. Such efforts will be essential to verify the global prevalence of dragon-king earthquakes and to refine our understanding of their generative mechanisms.

5.4 Dependence of dragon-king earthquakes on magnitude scales

The potential saturation of traditional magnitude scales such as local magnitude ($m_L$) and body-wave magnitude ($m_b$) for large earthquakes raises an important question regarding the reliability of dragon-king detection. However, a comparison of magnitude values for the 1975 Haicheng and 1976 Tangshan earthquakes suggests that saturation is not a dominant factor in these specific cases. For the Tangshan earthquake, the preferred moment magnitude ($M_W$) is 7.5–7.6, which is comparable to or slightly less than the reported surface-wave magnitude ($M_S$) of 7.8 and the local magnitude ($m_L$) of 7.9 (e.g., Chen et al., 1988). Similarly, for the Haicheng earthquake, $M_W$ is estimated at 7.0, consistent with or marginally less than the reported $M_S$ of 7.4. If significant saturation were affecting $m_L$ or $M_S$, we would expect $M_W$ to be substantially larger, which is not observed here. This indicates that the traditional magnitudes for these events have not been artificially capped, and thus the use of $m_L$ or $M_S$ does not systematically bias the mainshock's relative size in a way that would invalidate the DK test.

To quantitatively evaluate the sensitivity of dragon-king detection to the choice of magnitude scale, we computed *p*-values for the Haicheng and Tangshan mainshocks using both their traditionally reported magnitudes ($m_L$ 7.4 for Haicheng, $m_L$ 7.9 for Tangshan) and their preferred $M_W$ values (7.0 and 7.5, respectively). The results across density thresholds $f_{th}$ = 0.05, 0.15, and 0.25 for pre-mainshock, post-mainshock, and combined sequences are summarized in Table 1. Table 1 shows that, when the smaller $M_W$ values are used, the resulting *p*-values generally increase across most sequence types and density thresholds. For instance, for the Haicheng mainshock in the combined sequences, the *p*-value increases from approximately 0.047–0.061 when using $m_L$ 7.4 to 0.119–0.155 when using $M_W$ 7.0. A similar pattern is observed for the Tangshan mainshock. This systematic shift indicates that adopting a smaller mainshock magnitude makes the event appear less extreme relative to the GR distribution, thereby increasing the probability of accepting the null hypothesis (that it is not an dragon-king).

**Table 1.** *p*-values of the Haicheng and Tangshan mainshocks under different density thresholds $f_{th}$ and magnitude scales for pre-mainshock, post-mainshock, and combined sequences.

| **Mainshock** | ***m*** | $f_{th}$ **= 0.05** | | | $f_{th}$ **= 0.15** | | | $f_{th}$ **= 0.25** | | |
|---|---|---|---|---|---|---|---|---|---|---|
| | | **Pre-** | **Post-** | **All** | **Pre-** | **Post-** | **All** | **Pre-** | **Post-** | **All** |
| **Haicheng** | 7.4 | 0.010 | 0.056 | 0.047 | 0.006 | 0.076 | 0.060 | 0.006 | 0.058 | 0.047 |
| | 7.0 | 0.014 | 0.150 | 0.122 | 0.009 | 0.154 | 0.155 | 0.013 | 0.110 | 0.119 |
| **Tangshan** | 7.9 | 0.053 | 0.030 | 0.032 | 0.043 | 0.031 | 0.034 | 0.039 | 0.019 | 0.023 |
| | 7.5 | 0.121 | 0.094 | 0.105 | 0.113 | 0.095 | 0.104 | 0.141 | 0.064 | 0.070 |

Thus, although adopting smaller $M_W$ values weakens the signal, the analyses based on the traditional magnitudes still produce low *p*-values that robustly support a dragon-king interpretation. Furthermore, given that $m_L$ is sensitive to high-frequency energy while $M_W$ measures the static seismic moment, maintaining consistency between the magnitude scale of the mainshock and that of the surrounding seismicity (typically dominated by small to moderate events reported as $m_L$ or $m_b$ in local catalogs) is crucial. For the Haicheng and Tangshan earthquakes, the use of $M_S$ or $m_L$ is therefore more appropriate for ensuring a consistent comparative basis than adopting $M_W$ for the mainshock alone. It is also important to acknowledge that the mixing of magnitude types is a common, nearly universal feature of earthquake catalogs. Our analysis confirms that the primary conclusion, the statistical distinguishability of these mainshocks, holds when a self-consistent scale is used, and the primary goal of establishing their order within the sequence is adequately met. This investigation underscores the importance of explicit magnitude scale consideration in future dragon-king studies to ensure robust and reproducible results.

## 6 Conclusions

The present study has proposed a systematic framework for testing the dragon-king earthquake hypothesis through rigorous spatial clustering and statistical analysis. Defining dragon-kings as outliers driven by distinct physical mechanisms, we have first shown how to delineate coherent seismic regions using kernel density estimation (KDE) to avoid arbitrary boundaries and ensure geophysical consistency. For each identified cluster, we examined the tail of the frequency-magnitude distribution (FMD) to identify anomalous events. Candidate dragon-kings have been evaluated via robust statistical tests, primarily the max-robust-sum (MRS) test with inward sequential testing. We applied the framework to the seismicity associated with two well-known events: the $m_L$ 7.4 ($M_W$ 7.0-7.3) Haicheng earthquake (February 4, 1975) and the $m_L$ 7.9 ($M_W$ 7.5-7.6) Tangshan earthquake (July 28, 1976). In each case, we have calculated the $p$-value, defined as the probability that the observed statistic $T_{j,N-r}^{MRS}$ of the MRS test could be generated by the null distribution. We found that the Haicheng earthquake is strongly identified as a dragon-king event in its pre-mainshock sequence, with $p$-values ranging from 0.003 to 0.01 across a stable range of KDE density thresholds ($f_{th}$). Its post-mainshock and combined sequences yield slightly higher $p$-values, ranging from 0.04 to 0.09, but still remain below the conventional significance threshold of 0.1. The Tangshan earthquake also exhibits statistically significant dragon-king characteristics in both its pre-mainshock and post-mainshock sequences, with $p$-values ranging from 0.02 to 0.05. However, compared with Haicheng, the pre-mainshock dragon-king signal of Tangshan is less pronounced, suggesting that the two earthquakes may represent different temporal expressions of dragon-king behavior: Haicheng shows a clearer pre-mainshock signature, whereas Tangshan exhibits a weaker but still significant pre-mainshock signal and a comparably strong post-mainshock signature. This weaker and less clearly expressed pre-mainshock signal may have been one contributing factor, among many others, to the absence of an authoritative short-term warning in 1976. Although some precursory phenomena were reported for the Tangshan earthquake (see reviews and references cited in Mearns and Sornette, 2021a; Sornette et al., 2021), the available signals were not transformed into as compelling or coherent a statistical-physical pattern as in the Haicheng case, possibly reflecting not only data and scientific limitations but also the institutional, organisational, and political conditions under which the evidence was assessed and acted upon.

This comparison suggests that the dragon-king property is not simply an intrinsic binary label, but also depends on the temporal window and reference population used to construct the null hypothesis. For short-term forecasting, the pre-mainshock sequence is the most relevant window, because it tests whether the impending mainshock already stands out relative to the seismicity available before rupture. In this respect, Haicheng provides a clearer case of a precursor-rich dragon-king process, whereas Tangshan shows a significant but weaker pre-mainshock signal. By contrast, post-mainshock and combined sequences are more informative for characterizing the full rupture process, post-seismic relaxation, and long-term hazard implications. Thus, the dragon-king framework should be interpreted as a conditional and probabilistic diagnostic rather than as a deterministic prediction tool. Finally, while the statistical anomalies we identify are consistent with distinct physical mechanisms, we caution that such anomalies are not definitive proof of those mechanisms; they should be interpreted as strong diagnostic signals that, together with geophysical evidence, support the dragon-king hypothesis. Overall, the proposed framework provides a robust, transparent, and statistically grounded approach for identifying dragon-king earthquakes, with potential applications extending to other regions and types of extreme geophysical events.

It should also be emphasized that the pre-mainshock results remain constrained by the observational quality and limited sample size of the 1970s catalog, especially for Tangshan. A small number of pre-mainshock events may bias the estimation of the GR background and the construction of the null hypothesis, making $p$-values more sensitive to catalog completeness and threshold choices. Nevertheless, the Tangshan pre-mainshock $p$-values remain within the statistically significant range, supporting its classification as a dragon-king event while indicating a less robust short-term precursor signal than Haicheng. After the mainshocks, Tangshan displays lower $p$-values than Haicheng, suggesting a stronger post-mainshock dragon-king expression. Taken together, these contrasting properties indicate that some extreme earthquakes may exhibit clear pre-mainshock dragon-king signatures, whereas others may show more diffuse precursory behavior and reveal their full outlier character more strongly through rupture complexity and post-mainshock evolution. This underscores the importance of a statistically rigorous, cluster-based dragon-king framework for conditional probabilistic forecasting and long-term seismic hazard assessment.

Building on our findings, we outline a pathway for integrating the dragon-king framework into practical seismic risk mitigation. The methodology presented, specifically, the combination of KDE-based clustering and the sequentially updatable MRS test, is inherently suitable for operational deployment. It can be incorporated into real-time seismic monitoring systems to automatically identify emerging outlier foreshock sequences, thereby providing a novel, quantitative tool for short-term risk assessment. This practical application mirrors our successful implementation of the dragon-king framework for landslide prediction (Lei et al., 2023). Beyond automated algorithms, the core tenet of the dragon-king theory emphasizes dynamic risk management of extreme events through the monitoring and analysis of potential precursory signals (Sornette, 2009; Sornette and Ouillon, 2012). It is this philosophy that drives our ongoing collaborative initiative for Global Earthquake Forecasting System (GEFS; Sornette et al., 2021; Mignan et al., 2021). This interdisciplinary initiative aims to synthesize diverse geophysical precursors and conduct truly prospective, forward-testing prediction experiments (e.g., AoyuX earthquake predictability research system for Chinese mainland; Li et al., 2025c), ultimately translating the dragon-king concept from a theoretical model into a actionable strategy for seismic hazard preparedness.

**Acknowledgments**

The authors would like to thank Dr. Yang Zang from the China Earthquake Networks Center for his assistance in accessing the Chinese earthquake catalog data, and Prof. Yu Feng for providing the two induced seismicity datasets analyzed in this study. We are also deeply indebted to Dr. Davide Zaccagnino, Prof. Jiancang Zhuang , Prof. Zhongliang Wu, Prof. Qifu Chen, and Prof. Changsheng Jiang for their valuable suggestions. We sincerely thank the two anonymous reviewers for their insightful comments and constructive suggestions, which have greatly enhanced the quality of this manuscript. Jiawei Li wishes to express his deepest gratitude to his grandmother, Mrs. Zihua Li (1943-2026), whose steadfast support and unwavering belief in him have been a constant source of strength throughout the formative years of his research career. This work is partially supported by the Guangdong Basic and Applied Basic Research Foundation (Grant No. 2024A1515011568), Shenzhen Science and Technology Innovation Commission (Grant no. GJHZ20210705141805017), and the Center for Computational Science and Engineering at Southern University of Science and Technology.

**Open Research**

The Japan catalogs are available from the NIED (National Research Institute for Earth Science and Disaster Resilience) Data Management Center (https://www.fnet.bosai.go.jp/event/search.php?LANG=en). The catalog data for the Pohang, Korea earthquake are available from Woo et al. (2019), Yeo et al. (2020) and Kim et al. (2020). The Carbfix, Iceland earthquake catalog data are available from IS EPOS (2018). The Chinese data are acquired from the China Earthquake Networks Center (CENC) through the internal link provided by the Earthquake Cataloging System at China Earthquake Administration, available at http://10.5.160.18/console/index.action (last accessed: December 18, 2023), with a Digital Object Identifier (DOI) of 10.11998/SeisDmc/SN. This data is not publicly available; it can be requested from the CENC. The earthquake catalog for the study region, covering events with magnitudes ≥ 0 from the year 294 to 2024 (excluding hour-minute-second information and retaining only one decimal place for location precision), is available in the "Data" folder at https://github.com/lijiawei098/SeisDK_v1. The MATLAB codes implementing the dragon-king testing algorithm developed in this study are also openly accessible in the same GitHub repository.